\pdfoutput=1

\documentclass[review]{elsarticle}

\usepackage{lineno,hyperref}
\usepackage{amsmath}
\usepackage{arydshln}
\usepackage{graphicx}
\usepackage{booktabs}
\usepackage{threeparttable}
\usepackage{subfigure}
\modulolinenumbers[5]

\journal{Journal of Computational Physics}

\begin{document}

\begin{frontmatter}

\title{Stable, entropy-pressure compatible subsonic Riemann boundary condition for embedded DG compressible flow simulations}

\author[address1]{Ganlin Lyu}
\author[address2]{Chao Chen}
\author[address2]{Xi Du}
\author[address1]{Spencer J. Sherwin\corref{mycorrespondingauthor}}
\cortext[mycorrespondingauthor]{Corresponding author}
\ead{s.sherwin@imperial.ac.uk}

\address[address1]{Department of Aeronautics, Imperial College London, United Kingdom}
\address[address2]{Beijing Aircraft Technology Research Institute of COMAC, Beijing, 102211, China}


\begin{abstract}

One approach to reducing the computational cost of simulating transitional compressible boundary layer flow is to adopt a near body reduced domain with boundary conditions enforced to be compatible with a computationally cheaper three-dimensional RANS simulation. In such an approach it is desirable to enforce a consistent pressure distribution which is not typically the case when using the standard Riemann inflow boundary conditions. We therefore revisit the Riemann problem adopted in many DG based high fidelity formulations. 

Through analysis of the one-dimensional linearised Euler equations it is demonstrated that maintaining entropy compatibility with the RANS simulation is important for a stable solution. It is also necessary to maintain the invariant for the Riemann outflow boundary condition in a subsonic flow leaving one condition that can be imposed at the inflow boundary. Therefore the entropy-pressure enforcement is the only stable boundary condition to enforce a known pressure distribution. We further demonstrate that all the entropy compatible inflow Riemann boundary conditions are stable providing the invariant compatible Riemann outflow boundary condition is also respected. 

Although the entropy-pressure compatible Riemann inflow boundary condition is stable from the one-dimensional analysis, two-dimensional tests highlight divergence in the inviscid problem and neutrally stable wiggles in the velocity fields in viscous simulations around the stagnation point. A two-dimensional analysis about a non-uniform baseflow assumption provides insight into this stability issue (ill-posedness) and motivate the use of a mix of inflow boundary conditions in this region of the flow. 

As a validation we apply the proposed boundary conditions to a reduced domain of a wing section normal to the leading-edge of the CRM-NLF model taken out of a full three-dimensional RANS simulation at Mach 0.86 and a Reynolds number of 8.5 million. The results show that the entropy-pressure compatible Riemann inflow boundary condition leads to a good agreement in pressure distribution.
\end{abstract}


\end{frontmatter}

\linenumbers

\section{Introduction}
\label{introduction}
Boundary layer transition and the following turbulent boundary layer for external flows are of particular interest in the aerospace sector since a good aerodynamic design of a vehicle relies heavily on the correct prediction of transition position and turbulence features. In the transitional process, tiny vortices inside the boundary layer are generated, stretched, and finally break down, forming turbulence. To numerically study these fine evolutions, high-fidelity methods such as implicit large-eddy simulation (LES) and direct numerical simulation (DNS) are more regularly being adopted. However, limited by current computational resources, full-scale simulations at realistic Reynolds numbers ($\sim 10^7$) are not currently available, even at the most optimistic scenario predicted by Moore's law \cite{mengaldo2021industry}. As a result, current efforts into simulating transitional mechanisms often use high fidelity simulation in a near-body, reduced domain which is embedded in a outer domain encompassing the whole geometry where a lower fidelity model is applied. (See \cite{tempelmann2012swept} \cite{cooke2019destabilisation,cooke2020modelling} \cite{wenzel2019self,wenzel2019dns} \cite{dadfar2018control} \cite{hosseini2016direct}.)

Unlike in a large domain where the inflow and outflow boundaries are far away from the geometry of interest so that uniform freestream values can be enforced for the boundary conditions, simple boundary conditions are not available on the reduced domain. Hence, the distributions of the solution fields need to be specified in advance. For example, Wenzel \cite{wenzel2019dns} studied the high subsonic turbulent boundary layer in a rectangular reduced domain on the wing of Boeing 787 using DNS. To enforce the freestream boundary condition, a modelled velocity distribution was provided through analytical expression, and the temperature and pressure distributions were calculated according to the one-dimensional isentropic relation. An alternative approach at the boundaries is to use the results from a Reynolds-Averaged Navier–Stokes (RANS) simulation, which is carried in a larger or full domain. The RANS result can then be interpolated along the boundary of the reduced domain to provide appropriate distributions, e.g. Cooke \cite{cooke2020modelling}, Tempelmann \cite{tempelmann2012swept}, and other references mentioned above. In these studies, the velocity distributions at the inflow boundary (of the C-type mesh) are interpolated. Since the inflow boundary is located upstream of the geometry, the fields are not typically influenced by the turbulence model in RANS. Compared with the distribution modelling which approximates field distributions along the boundary by analytical expressions, interpolating the distributions from a RANS simulation is more flexible. This enables the cases with more complex geometries to be studied, where the distributions are not typically well described by analytical expressions.

Having obtained the desired  distributions on the boundary, it then needs to be enforced on the  boundary of the reduced domain through an appropriate implementation. The types of boundary conditions are usually different for incompressible flows and compressible flows. The governing equations for incompressible flows are often solved with strong boundary conditions which enforce the velocity and pressure through Dirichlet or Neumann boundary conditions. Therefore, the desired distributions of data from the outer simulations can be directly enforced. On the other hand, weak boundary conditions enforcement, for example using a Riemann solution to infer a desired numerical flux, are often adopted in compressible flow solvers to achieve better accuracy \cite{abbas2010weak} and convergence \cite{eliasson2009influence,nordstrom2012weak}. Compared with the strong enforcement, the flow quantity distributions are usually not directly set in the weak enforcement.

In addition, in the compressible flow regime, subsonic flows and supersonic flows behave differently due to the different sign of the hyperbolic characteristics, and therefore, different boundary conditions have to be specified. The most significant consequence lies in the number of conditions to well-pose the problem \cite{nordstrom2005well}. In a Discontinuous Galerkin (DG) and Riemann-based solver, and for the inviscid fluxes of the governing equations, two conditions can be imposed in the inwards normal direction to the boundary for a subsonic inflow while three conditions can be imposed for a supersonic inflow. The choice of conditions naturally leads to the different quantities being enforced on the subsonic inflow boundary. A widely adopted subsonic inflow boundary condition in a DG solver is to enforce the incoming Riemann invariant and entropy, leading to a non-reflecting boundary. However, this does not directly enforce a compatible pressure condition with the outer RANS simulation. This incompatibility in the pressure distribution is undesirable in boundary layer transition prediction since the streamwise pressure gradient has a significant influence on the development of a transitional boundary layer flow. For example the Tollmien–Schlichting waves are stabilized by a favorable pressure gradient and destabilized by an adverse pressure gradient while both pressure gradients destabilize the crossflow waves \cite{reed1996linear,saric2003stability}. Further the pressure load is usually well captured by the Euler or RANS simulation (at least for lift prediction) and the pressure distribution does not typically vary much over the boundary layer. The pressure distribution is therefore considered a reliable quantity of interest from the lower fidelity models and is an important property to be maintained in the reduced domain.

To achieve the pressure compatibility in the reduced domain, the corresponding subsonic inflow boundary condition needs to be constructed. However, although this construction can follow several approaches, not all of them are appropriate. The ideal boundary conditions enable the numerical simulation to converge in a stable way, inferring that the partial differential equation problem is well-posed and the numerical discretisation is stable with the adopted boundary conditions.

For the stability analysis, Giles \cite{giles1983eigenmode} and Darmofal et al. \cite{darmofal2000eigenmode} have developed one-dimensional (1D) linear analysis (or eigenmode analysis) to evaluate decay rates of the disturbances. Since the decay rates depend on the adopted boundary conditions, the performance of boundary conditions are therefore obtained. This method can also be used to check the well-posedness of the boundary condition enforcement. Nordstr\"{o}m et al. \cite{nordstrom2005well,nordstrom2017roadmap} used the energy method to analyze the linearized system in a control volume. This method can be used to understand the number of conditions to make the problem well-posed as well as examine suitability of certain combination of boundary conditions. These works provide insightful analysis and set up a useful framework, which are wildly used to establish boundary conditions for compressible flow simulations.

As for the boundary condition stability analysis of DG approximations, it is less studied compared with those based on the conventional finite difference discretisation. Hindenlang et al. \cite{hindenlang2020stability} examined the linear stability of the wall boundary condition for the inviscid flows using the sufficient condition developed by Gassner et al. \cite{gassner2018br1}. This study focused on the wall boundary condition while reasonably assumed the conditions on other boundaries are stable. However, since it is the combination of boundary conditions that determines the evolution of the system, the satisfaction of this condition on individual boundaries does not necessarily guarantee a non-singular system. To obtain a specific analysis within the conventional method, where the disturbance is introduced internally, a uniform baseflow is sometimes needed to compute the integration over the boundaries of the control volume. This can limit its application to problems with large gradient, and therefore an extension of this methodology is required.

In this paper we therefore revisit the Riemann problem to see how best to enforce pressure compatibility at the subsonic inflow boundary. In Section \ref{sec::Linearized_DG_for_Euler1D}, one-dimensional (1D) Euler equations are linearized in a DG element, and the systems for the inflow boundary conditions to be stable are derived from two perspectives. An entropy-pressure compatible inflow boundary condition is selected from the full range of possibilities in Section \ref{sec::stabilityAnalysis} . Section \ref{sec::illPosedness_2D} extends the current analysis to multi-dimensions without assuming an uniform baseflow assumption. This method is then used to understand the instability issue in adopting a stale 1D entropy-pressure inflow in the region of a stagnation point flow. Finally, a transonic flow in a reduced domain taken out of a full three-dimensional (3D) RANS simulation is computed for validation.

\section{Linearized DG approximation for 1D Euler equations}
\label{sec::Linearized_DG_for_Euler1D}
To analyze the applicability of weak boundary conditions, we now consider the stability of the linearized Euler system subject to a discontinuous Galerkin (DG) discretisation. A common practice for DG approximations 
of the compressible flow equations is to impose  boundary conditions  using one-dimensional characteristic line theory. We therefore follow this practice and derive the linearized DG approximation for 1D Euler equations, whose conservative form reads
\begin{equation}
  \frac{\partial \mathbf{Q}}{\partial t}
+ \frac{\partial \mathbf{F(\mathbf{Q})}}{\partial x} = \mathbf{0}
\label{eq::Euler1D_differenial}
\end{equation}
where $\mathbf{Q}$ is the state vector of conservative variables, $\mathbf{F}$ is vector for inviscid flux, that is
\begin{equation}
  \mathbf{Q} =\left(
    \begin{array}{c}
      \rho \\ \rho u \\ E
    \end{array} \right), \hspace{0.5 cm}
  \mathbf{F} =\left(
    \begin{array}{c}
      \rho u \\ p + \rho u^2 \\ u (E + p)
    \end{array} \right). 
\end{equation}
In the above  $\rho$ is the density, $u$ is the velocity in the $x$-direction, $p$ is the pressure and $E$ is the total energy, which can be expressed under the assumption of a perfect gas law as
\[
  E = \frac{1}{\gamma - 1} p + \frac{1}{2} \rho u^2,
\]
where $\gamma$ is the specific heat ratio.


\begin{figure}[htbp]
  \centering
  \includegraphics[width=7cm]{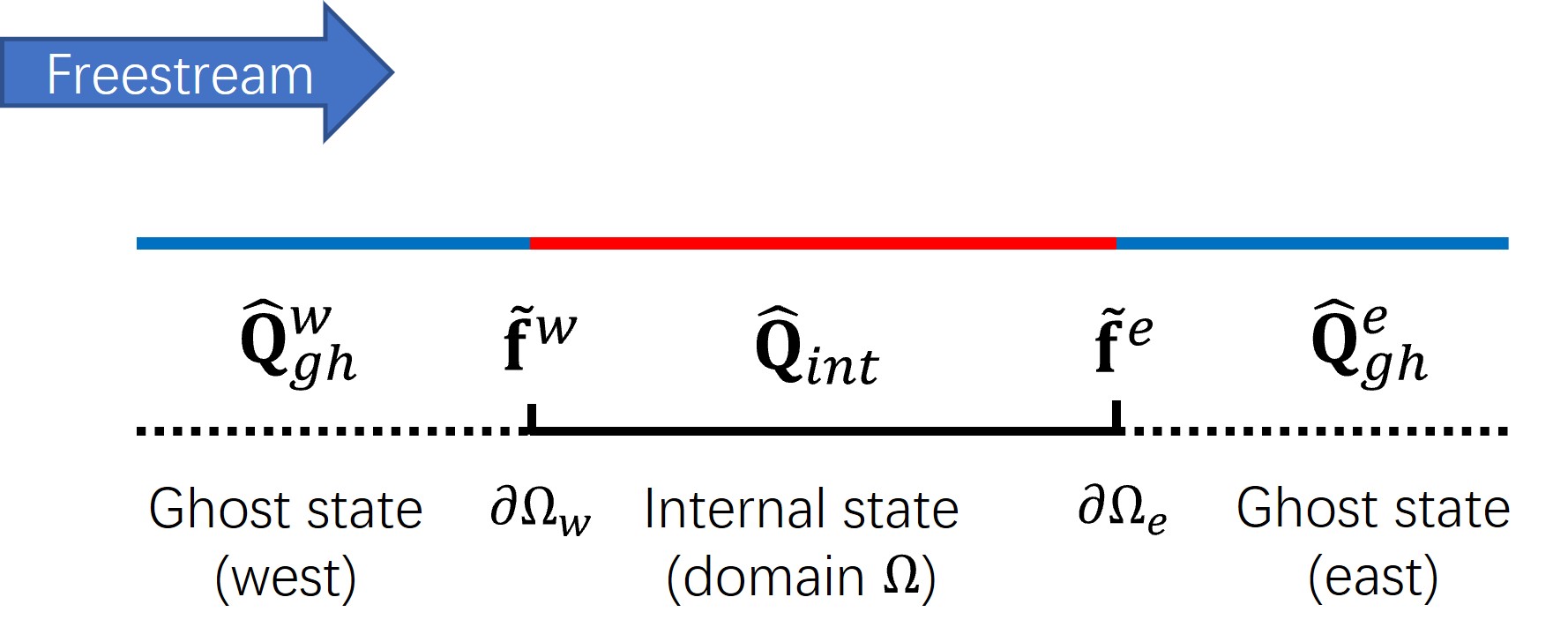} 
  \caption{1D solution domain and associated conservative states. The red region represents the inner state and blue regions represent the ghost state on each side where the solution domain boundaries are provided.}
  \label{Fig_1D_Q}
\end{figure}

In the DG approach, the approximation in each  element are independent and boundary conditions are enforced weakly in each element in the form of a numerical flux. Therefore, to derive the smallest linear system for stability analysis which  preserves all the necessary boundary conditions, we consider the domain as consisting of a single element $\Omega$, as shown in Fig. \ref{Fig_1D_Q}. Eq. (\ref{eq::Euler1D_differenial}) is first multiplied by a test function $\phi$ and then integrated in the domain by part, which gives
\begin{equation}
\label{eq::Euler1D_DG_1}
  \int_{\Omega} \phi \frac{\partial\mathbf{Q}}{\partial t}  dx  + 
 \left [  \phi \mathbf{F}\cdot \mathbf{n} \right ]_{\partial \Omega} - \int_{\Omega} \nabla \phi \cdot \mathbf{F} = 0
\end{equation} 


For simplicity we next consider a piecewise  constant approximation in the element
\begin{equation}
\label{eq::pieceConstantApproximation}
  \mathbf{Q} \simeq \hat{\mathbf{Q}}(t) \phi(x), \hspace{0.5 cm}
  \phi(x) = \left\{ 
  \begin{aligned} 
    & 1, \quad x \in \Omega  \\
    & 0, \quad x \notin \Omega 
  \end{aligned} \right.
\end{equation} 
Imposing this piecewise constant approximation to Eq. (\ref{eq::Euler1D_DG_1}) allows us to analytically evaluate the first term and directly sets the third term to zero. Finally  replacing the boundary flux $\mathbf{F}$ by a numerical flux  denoted by $\tilde{\mathbf{f}}$, Eq. (\ref{eq::Euler1D_DG_1}) becomes
\begin{equation}
\label{eq::Euler1D_DG_2}
  \frac{d\hat{\mathbf{Q}}}{d t} \Delta x = \tilde{\mathbf{f}}^{w} - \tilde{\mathbf{f}}^{e} 
\end{equation}
where $\Delta x = x^e-x^w$ is the length of the 1D domain and the superscript $w$ and $e$ denote the west and east boundaries of the element. The numerical flux is typically computed through a Riemann solver, and takes the form
\begin{equation}
  \tilde{\mathbf{f}} = \tilde{\mathbf{f}}(\hat{\mathbf{Q}}_{int},\hat{\mathbf{Q}}_{ext})
\end{equation}
where at the interface between two adjacent elements $\hat{\mathbf{Q}}_{int}$ and $\hat{\mathbf{Q}}_{ext}$ represent the internal state and external states. For the numerical flux at the solution domain boundary, where there is no adjacent element, a ghost state  $\hat{\mathbf{Q}}_{gh}$ is introduced instead of the external state  so that we have
\begin{equation}
  \tilde{\mathbf{f}} = \tilde{\mathbf{f}}(\hat{\mathbf{Q}}_{gh},\hat{\mathbf{Q}}_{int})
\end{equation}
In the standard use of a Riemann boundary condition the external or ghost state is independent of the interior state and the solution of the Riemann solver will provide a known boundary state $\hat{\mathbf{Q}}_b$, i.e.
\[
\hat{\mathbf{Q}}_b(\hat{\mathbf{Q}}_{gh},\hat{\mathbf{Q}}_{int}). \]
In this circumstances $\hat{\mathbf{Q}}_b$ is not known explicitly but rather inferred through  $\hat{\mathbf{Q}}_{gh}$ and  $\hat{\mathbf{Q}}_{int}$ and the Riemann solution procedure \cite{mengaldo2014guide}. Nevertheless it is also possible to explicitly impose as many conditions as there are incoming characteristics on the domain boundary,  for example a prescribed freestream quantity such as pressure or a zero normal velocity at an inviscid wall. For this situation we  introduce 
 $\hat{\mathbf{Q}}_{ref}$ to denote an external reference state which when provided as input (or the ghost state) to the Riemann solver in combination with the internal state it enforces the desired condition on $\hat{\mathbf{Q}}_b$ and subsequently $\tilde{f}(\hat{\mathbf{Q}}_b)$. However in this case the reference state $\hat{\mathbf{Q}}_{ref}$ is now dependent on $\hat{\mathbf{Q}}_{int}$ and the desired conditions we wish to impose such that
 
\begin{equation}
\begin{aligned}
 \tilde{\mathbf{f}} & = \tilde{\mathbf{f}}(\hat{\mathbf{Q}}_{ref}(\hat{\mathbf{Q}}_{int}), \hat{\mathbf{Q}}_{int}) \\ 
  & = \tilde{\mathbf{f}}(\hat{\mathbf{Q}}_{ref},\hat{\mathbf{Q}}_{int}) 
\end{aligned}
\end{equation}

\begin{figure}[htbp]
  \centering
  \includegraphics[width=7cm]{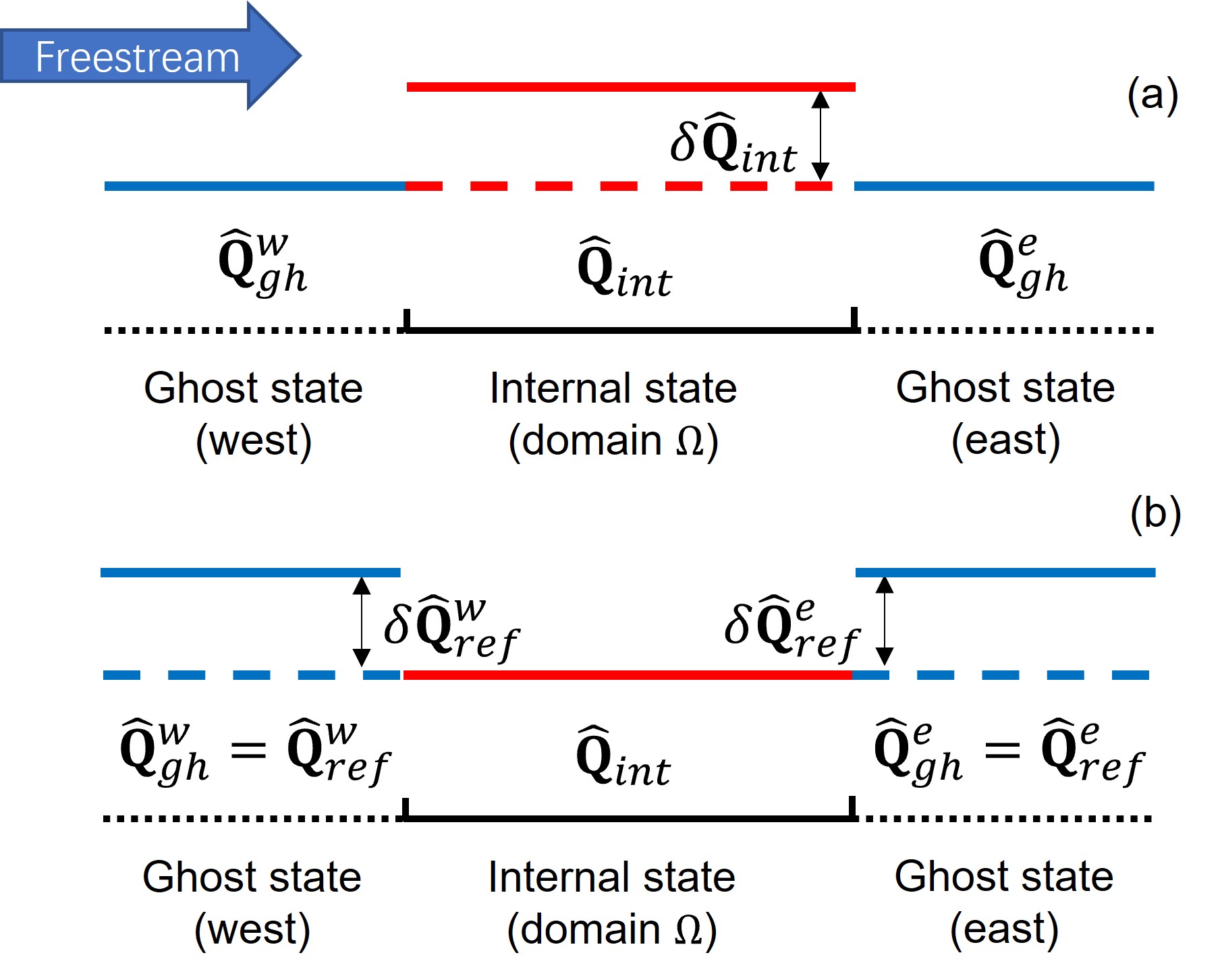} 
  \caption{Steady state and perturbation in 1D cases. (a) Perturbation on the inner state; (b) perturbation on the Reference state. The dashed lines represent the steady state.}
  \label{Fig_1D_Q_ptub}
\end{figure}

Since the numerical flux is constructed with the reference state and the internal state, the linearized system for stability analysis can be generated by introducing disturbances on both states, as is shown in Fig. \ref{Fig_1D_Q_ptub}. The linearized form of Eq. (\ref{eq::Euler1D_DG_2})  (see \ref{sec::appendix_A} for full details), is given by
\begin{equation}
\label{eq::LinStability1D_Q}
  \frac{d \left(\delta \hat{\mathbf{Q}}_{int} \right)}{d t} \Delta x = \left( 
  \frac{\partial \tilde{\mathbf{f}}^w}{\partial \hat{\mathbf{Q}}_{int}} - \frac{\partial \tilde{\mathbf{f}}^e}{\partial \hat{\mathbf{Q}}_{int}} \right)  \delta \hat{\mathbf{Q}}_{int} + \left( 
  \frac{\partial \tilde{\mathbf{f}}^w}{\partial \hat{\mathbf{Q}}_{ref}^w}  - \frac{\partial \tilde{\mathbf{f}}^e}{\partial \hat{\mathbf{Q}}_{ref}^e} \right) \delta \hat{\mathbf{Q}}_{ref}
\end{equation}
where $\delta \hat{\mathbf{Q}}_{int}$ is the time-dependent disturbance on the internal state and $\delta \hat{\mathbf{Q}}_{ref}$ is the constant disturbance on the reference state. (The disturbance on the west boundary is assumed equals to that on the east boundary.)

Although the independent variables in  Eq. (\ref{eq::LinStability1D_Q}) can be in any form for the 1D analysis it is convenient to use the characteristic variables which are 
\begin{equation}
\label{eq::characteristicVars}
  \mathbf{U} =\left(
  \begin{array}{c}
    R^+ \\  R^0 \\ R^-
  \end{array} \right) = \left(
  \begin{array}{c}
    u+\frac{2}{\gamma-1} c \\  S  \\ u-\frac{2}{\gamma-1} c
  \end{array} \right)
\end{equation}
where $c=\sqrt{\gamma p/\rho}$ is the speed of sound, $S$, which is the measure of entropy and takes the form
\[
  S = \frac{p}{\rho^\gamma}
\]

By assuming the baseflow is steady and therefore uniform because of one-dimensionality, the 1D linearized system in characteristic variables takes the form
\begin{equation}
\label{eq::LinStability1D_U}
  \frac{d}{d t} \left(\delta \hat{\mathbf{U}}_{int} \right) =
  \frac{1}{\Delta x} \mathbf{C}_{int} \delta \hat{\mathbf{U}}_{int} + \frac{1}{\Delta x} \mathbf{C}_{ref} \delta \hat{\mathbf{U}}_{ref} 
\end{equation}
where the coefficient matrices are defined as 
\begin{equation}
\label{eq::CoefMatrix_int}
  \mathbf{C}_{int} = 
  \left(\frac{\partial \hat{\mathbf{Q}}}{\partial \hat{\mathbf{U}}} \right)^{-1}
  \frac{\partial \tilde{\mathbf{f}}}{\partial \hat{\mathbf{Q}}}
  \frac{\partial \hat{\mathbf{Q}}}{\partial \hat{\mathbf{U}}}
  \left(
  \frac{\partial \hat{\mathbf{U}}_b^w}{\partial \hat{\mathbf{U}}_{int}} - \frac{\partial \hat{\mathbf{U}}_b^e}{\partial \hat{\mathbf{U}}_{int}}
  \right) = \mathbf{C}_{1}\mathbf{C}_{2,int},
\end{equation}
\begin{equation}
\label{eq::CoefMatrix_ref}
  \mathbf{C}_{ref} = 
  \left(\frac{\partial \hat{\mathbf{Q}}}{\partial \hat{\mathbf{U}}} \right)^{-1}
  \frac{\partial \tilde{\mathbf{f}}}{\partial \hat{\mathbf{Q}}}
  \frac{\partial \hat{\mathbf{Q}}}{\partial \hat{\mathbf{U}}}
  \left(
  \frac{\partial \hat{\mathbf{U}}_b^w}{\partial \hat{\mathbf{U}}_{ref}^w} - \frac{\partial \hat{\mathbf{U}}_b^e}{\partial \hat{\mathbf{U}}_{ref}^e} \right) = \mathbf{C}_{1}\mathbf{C}_{2,ref}.
\end{equation}
In the above, $\mathbf{C}_{1}$  relates to the uniform baseflow quantities and can be further evaluated as: 
\begin{equation}
\label{eq::CoefMatrix_1}
  \mathbf{C}_1 = \left(\frac{\partial \hat{\mathbf{Q}}}{\partial \hat{\mathbf{U}}} \right)^{-1} \frac{\partial \tilde{\mathbf{f}}}{\partial \hat{\mathbf{Q}}}
  \frac{\partial \hat{\mathbf{Q}}}{\partial \hat{\mathbf{U}}} =\left(
  \begin{array}{c c c}
    c+u & -\frac{c^2}{\gamma(\gamma-1)S} & 0 \\
      0 &         u                      & 0 \\
      0 & -\frac{c^2}{\gamma(\gamma-1)S} & u-c
  \end{array} \right)
\end{equation}
which is similar to the Jacobian matrix $\partial \tilde{\mathbf{f}}/\partial \hat{\mathbf{Q}}$, and therefore the eigenvalues are the well-known $u+c$, $u$, and $u-c$. The remaining  coefficient matrices are
\begin{equation}
\label{eq::CoefMatrix_2_int}
  \mathbf{C}_{2,int} = 
  \frac{\partial \hat{\mathbf{U}}_b^w}{\partial \hat{\mathbf{U}}_{int}} -
  \frac{\partial \hat{\mathbf{U}}_b^e}{\partial \hat{\mathbf{U}}_{int}}
\end{equation}
\begin{equation}
\label{eq::CoefMatrix_2_ref}
  \mathbf{C}_{2,ref} = 
  \frac{\partial \hat{\mathbf{U}}_b^w}{\partial \hat{\mathbf{U}}_{ref}^w} -
  \frac{\partial \hat{\mathbf{U}}_b^e}{\partial \hat{\mathbf{U}}_{ref}^e}
\end{equation}
which contain the normalized coefficients about how the boundary condition will respond to the disturbance of the internal and reference states, respectively.

For the coefficient matrix $\mathbf{C}_{int}$, we assume it can be diagonalised so that 
$$ \mathbf{C}_{int} = \mathbf{P}^{-1} \mathbf{\Lambda} \mathbf{P} $$
where $\mathbf{\Lambda} = diag\left( \lambda_i \right)$ ($i=1,2,3$), and $\lambda_i$ is its eigenvalue. If these eigenvalues are non-zero, the solution to Eq. (\ref{eq::LinStability1D_U}) can be written as
\begin{equation}
\label{eq::LinStability1D_U_solution}
  \mathbf{P} \delta\hat{\mathbf{U}}_{int} (t) = e^{\mathbf{\Lambda} t/\Delta x} \mathbf{P} \delta\hat{\mathbf{U}}_{int} (0) + \left(\mathbf{I} - e^{\mathbf{\Lambda} t/\Delta x}\right) \left( -\mathbf{P} \mathbf{C}_{int}^{-1} \mathbf{C}_{ref} \right)  \delta\hat{\mathbf{U}}_{ref}.
\end{equation}
For a stable solution to a well-posed problem, the initial disturbance on the internal state $\delta\hat{\mathbf{U}}_{int} (0)$ in Eq. (\ref{eq::LinStability1D_U_solution}) should decay. Physically we understand that the disturbance waves will leave the domain through the boundaries without generating a stronger reflection, meanwhile the disturbance on the reference state enters the domain. The solution should exponentially converge to the reference value of the boundary conditions. This result is achieved if, and only if, the following conditions are satisfied
\begin{equation}
\label{eq::LinStability1D_U_requirement}
  \left\{
  \begin{aligned}
  &  Re(\lambda_i) < 0, \hspace{0.3 cm} i=1,2,3\\
  &  \mathbf{C}_{ref} = - \mathbf{C}_{int}.
  \end{aligned}
  \right.
\end{equation}
The second condition is equivalent to a opposite relation for the second part of the coefficient matrices as 
\begin{equation}
\label{eq::LinStability1D_U_requirement_equivalent}
  \mathbf{C}_{2,ref} = - \mathbf{C}_{2,int}
\end{equation}
If this condition is not satisfied, the solution will converge to
\begin{equation}
  \delta\hat{\mathbf{U}}_{int} (t \rightarrow \infty) = -\mathbf{C}_{2,int}^{-1}\mathbf{C}_{2,ref} \delta\hat{\mathbf{U}}_{ref} \ne \delta\hat{\mathbf{U}}_{ref}
\end{equation}
This solution fails to track the reference value, and therefore is non well-posed.  

Another possible scenario is when $\mathbf{C}_{int}$ is singular so that some eigenvalue is zero, e.g. $\lambda_i=0$, the $i$-th component solution becomes
\begin{equation}
\label{eq::LinStability1D_U_solution_zero}
  \left(\mathbf{P}\delta\hat{\mathbf{U}}_{int}(t)\right)_i = \left(\mathbf{P}\delta\hat{\mathbf{U}}_{int}(0)\right)_i + 
  \left(\frac{1}{\Delta x}\mathbf{P} \mathbf{C}_{ref} \delta\hat{\mathbf{U}}_{ref}\right)_i t
\end{equation}
This solution will grow or decay with time monotonously and finally diverge unless the entries in the $i$-th row of $\mathbf{P}\mathbf{C}_{ref}$ are all zero, inferring that $\mathbf{C}_{ref}$ is also singular as $\mathbf{P}$ has full rank. Under this condition the component solution is neutrally stable, which is accepted from the energy point of view since the disturbance will not further grow and thus the solution is bounded. However, its evolution is unrelated to the boundary conditions through the reference value. Since  Eq. (\ref{eq::LinStability1D_U}) is linearized from the discrete approximation given by  Eq. (\ref{eq::Euler1D_DG_2}), the failure to track the reference value further indicates the approximation is ill-posed through the boundary condition enforcement. Or alternatively, this neutral stability violates the uniqueness requirement for a well-posed steady state boundary value problem because both the steady state solution and the solution with the eigenfunction of the zero eigenvalue superimposed also satisfy the boundary conditions. 

According to the above analysis, the two requirements to form a stable system are provided by Eq. (\ref{eq::LinStability1D_U_requirement}), based on the eigenvalue analysis of the coefficient matrices. We therefore next consider how the coefficient matrices are constructed for specific Riemann boundary conditions.

\subsection{Characteristic treatment for Riemann boundary conditions}
\label{sec::characteristicTreatment}


Having established the linear system, Eq. (\ref{eq::LinStability1D_U}), relating the stability of the internal solution to the boundary conditions, we now need to construct the second part of coefficient matrices, $\mathbf{C}_{2,int}$ and $\mathbf{C}_{2,ref}$. Before doing so we first need to outline the characteristic treatment for Riemann boundary conditions.

\begin{figure}[htbp]
\label{Fig_characteristics}
  \centering
  \includegraphics[width=6cm]{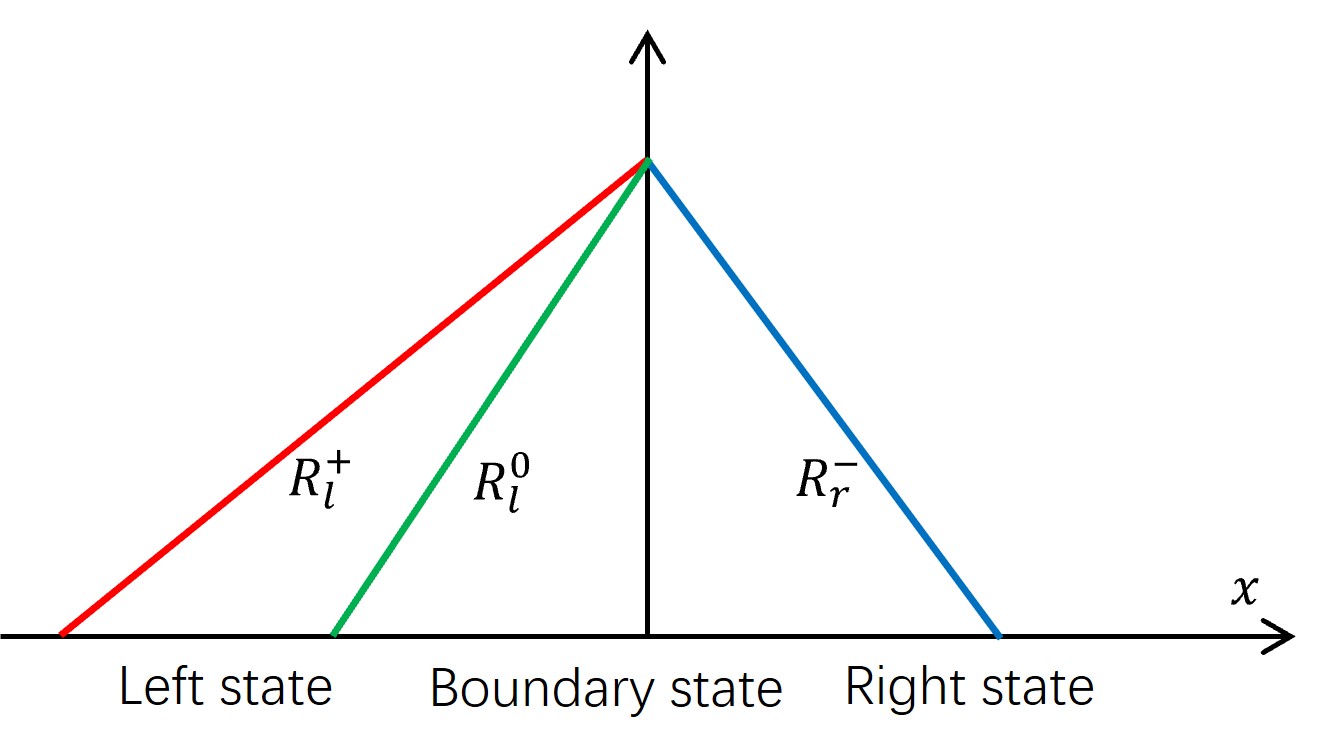} 
  \caption{Characteristic lines for a subsonic boundary.}
\end{figure}

To set up a Riemann boundary condition, a common practice is to construct the boundary state following the characteristic relations. Fig. \ref{Fig_characteristics} shows the characteristics (or characteristic curves) for the 1D Euler equations at a subsonic boundary ($x=0$). For a subsonic inflow, the left state can be considered as  the ghost state and two characteristics are pointing into the domain, indicating two conditions can be  enforced, while the third conditions is provided by the solution within the domain (right state) related to the outwards-pointing characteristic. Since the Riemann invariants, and entropy, are constant along the characteristics, the following relations have to be satisfied at the boundary
\begin{equation}
\label{eq::characteristic_relation}
  R_b^+ = R_l^+, \hspace{0.5 cm}
  R_b^0 = R_l^0, \hspace{0.5 cm}
  R_b^- = R_r^-
\end{equation}
where subscript $l$ and $r$ denote the left (ghost) state and right (internal) state, respectively. 

The simplest boundary condition one can enforce 
is to specify the reference state to be a given ghost state independent of the internal values (as is in Fig. \ref{Fig_1D_Q_ptub}) i..e 
\begin{equation}
  \hat{\mathbf{Q}}_{ref} = \hat{\mathbf{Q}}_{gh}
\end{equation}
which leads to the so called Riemann inflow \cite{darmofal2000eigenmode}. We will also refer to this boundary condition as an entropy-invariant compatible inflow in the current work since the entropy (as well as the other incoming Riemann invariant) are directly enforced on the boundary state by an exact Riemann solver. 

Since this inflow boundary condition only enforces entropy and the incoming Riemann invariant, there is no guarantee for pressure compatibility with the outer solution.  To complete the stability analysis, the boundary condition at the outflow also needs to be provided. In what follows we will enforce the invariant compatible condition at the outflow in all of the following cases. 
This inflow-outflow combination, denoted by superscript $SI$, leads to the following coefficient matrices
\begin{equation}
\label{eq::CoefMatrix_2_int_SI}
  \mathbf{C}_{2,int}^{SI} = 
  \frac{\partial \hat{\mathbf{U}}_b^w}{\partial \hat{\mathbf{U}}_{int}} -  \frac{\partial \hat{\mathbf{U}}_b^e}{\partial \hat{\mathbf{U}}_{int}} = \left(
  \begin{array}{c c c}
    0 & 0 & 0 \\
    0 & 0 & 0 \\
    0 & 0 & 1
 \end{array} \right) - \left(
  \begin{array}{c c c}
    1 & 0 & 0 \\
    0 & 1 & 0 \\
    0 & 0 & 0
 \end{array} \right) = \left(
  \begin{array}{c c c}
    -1 &  0 & 0 \\
     0 & -1 & 0 \\
     0 &  0 & 1
 \end{array} \right)
\end{equation}
\begin{equation}
\label{eq::CoefMatrix_2_ref_SI}
  \mathbf{C}_{2,ref}^{SI} = 
  \frac{\partial \hat{\mathbf{U}}_b^w}{\partial \hat{\mathbf{U}}_{ref}^w} -
  \frac{\partial \hat{\mathbf{U}}_b^e}{\partial \hat{\mathbf{U}}_{ref}^e} = \left(
  \begin{array}{c c c}
    1 & 0 & 0 \\
    0 & 1 & 0 \\
    0 & 0 & 0
 \end{array} \right) - \left(
  \begin{array}{c c c}
    0 & 0 & 0 \\
    0 & 0 & 0 \\
    0 & 0 & 1
 \end{array} \right) = \left(
  \begin{array}{c c c}
    1 & 0 &  0 \\
    0 & 1 &  0 \\
    0 & 0 & -1
 \end{array} \right)
\end{equation}

By substituting Eq.  (\ref{eq::CoefMatrix_2_int_SI}) and (\ref{eq::CoefMatrix_2_ref_SI}) into Eqs. (\ref{eq::CoefMatrix_int}) and (\ref{eq::CoefMatrix_ref}), the coefficient matrices becomes
\begin{equation}
\label{eq::CoefMatrix_int_SI}
  \mathbf{C}_{int}^{SI} = \mathbf{C}_1 \mathbf{C}_{2,int}^{SI} = \left(
  \begin{array}{c c c}
    -(c+u) & \frac{c^2}{\gamma(\gamma-1)S} & 0 \\
         0 &         -u                    & 0 \\
         0 & \frac{c^2}{\gamma(\gamma-1)S} & -(c-u)
  \end{array} \right) = - \mathbf{C}_{ref}^{SI}
\end{equation}
where the second condition in Eq. (\ref{eq::LinStability1D_U_requirement}) is directly satisfied. The eigenvalues of $\mathbf{C}_{int}^{SI}$ are $-(c+u)$, $-u$, and $-(c-u)$ while eigenvalues for $\mathbf{C}_{ref}^{SI}$ are the same but negative values. The physical interpretation of these eigenvalues is that when the internal state is disturbed, the disturbance can be decomposed onto the three waves, which leave the domain at the speeds $c+u$,$u$, and $c-u$. The minus signs represent the decaying feature of the disturbances. On the other hand, if the reference state is disturbed, the disturbance can also be projected onto the waves that come into the domain at the positive speed $c+u$, $u$, and $c-u$. 
Since the eigenvalues for $\mathbf{C}_{int}^{SI}$ are all negative, the first condition in Eq. (\ref{eq::LinStability1D_U_requirement}) is also satisfied. Therefore, the solution to the linearized system in Eq. (\ref{eq::LinStability1D_U}) is stable at the steady state using these boundary conditions.

We note from inspecting Eqs. (\ref{eq::CoefMatrix_2_int_SI}) and (\ref{eq::CoefMatrix_2_ref_SI}) that the following relations hold
\begin{equation}
\label{eq::CoefMatrix_2_relation_1}
  \frac{\partial \hat{\mathbf{U}}_b^w}{\partial \hat{\mathbf{U}}_{int}} + \frac{\partial \hat{\mathbf{U}}_b^w}{\partial \hat{\mathbf{U}}_{ref}^w} = \mathbf{I}, \hspace{0.5 cm}
  \frac{\partial \hat{\mathbf{U}}_b^e}{\partial \hat{\mathbf{U}}_{int}} + \frac{\partial \hat{\mathbf{U}}_b^e}{\partial \hat{\mathbf{U}}_{ref}^e} = \mathbf{I}
\end{equation}
\noindent and 
\begin{equation}
\label{eq::CoefMatrix_2_relation_2}
  \mathbf{C}_{2,int}^{SI} + \mathbf{C}_{2,ref}^{SI} = \mathbf{0}. 
\end{equation}
These relations are satisfied not only for the current boundary conditions but for all combinations including supersonic ones. This is because to construct the boundary state specific information should come from either internal state or reference state or both. The sum of the weight of contributions from both states should be one or zero depends on whether it is a diagonal entry, and form an identity matrix. Therefore the opposite condition in Eq. (\ref{eq::LinStability1D_U_requirement}) (or (\ref{eq::LinStability1D_U_requirement_equivalent})) is always satisfied by the Riemann boundary conditions, and the eigenvalues for $\mathbf{C}_{2,int}$ and $\mathbf{C}_{2,ref}$ are always equal and opposite numbers.

Although the opposite relation hold, we still distinguish $\mathbf{C}_{int}$ and $\mathbf{C}_{ref}$ since they are computed through different derivatives and this provides two different perspectives to examine if the boundary conditions enforcement leads to a stable system. In Eq. (\ref{eq::LinStability1D_U}) if the disturbance on the reference state is not considered, it recovers the homogeneous dynamical system for conventional stability analysis
\begin{equation}
\label{eq::LinStability1D_U_int}
  \frac{d}{d t} \left(\delta \hat{\mathbf{U}}_{int} \right) =
  \frac{1}{\Delta x} \mathbf{C}_{int} \delta \hat{\mathbf{U}}_{int}
\end{equation}
The stability of this system can be analyzed directly through the eigenvalues of $\mathbf{C}_{int}$. Alternatively, we can also examine the stability by constructing $\mathbf{C}_{ref}$ and using the opposite relation. 

A special case that is worth further discussion is when one or more eigenvalues in $\mathbf{C}_{2,int}$ (or equivalently in $\mathbf{C}_{2,ref}$) are zero. According to the analysis in the previous section, in this case the solution in Eq. (\ref{eq::LinStability1D_U_solution_zero}) is neutrally stable but the original problem is ill-posed. The zero eigenvalues usually occur when the same reference quantity is enforced at both inflow and outflow boundaries. This quantity could then be over constrained and there will be some other quantify lacking a constraint. In the construction of the coefficient matrices, this appears in the form of cancellation of entries, leading to rank-deficiency. Further similar to this zero eigenvalue case is the situation where some eigenvalues are purely imaginary. The difference lies in that the zero eigenvalues arise from insufficient boundary conditions to a well-pose problem while the purely imaginary eigenvalues indicates that sufficient boundary conditions are enforced but in a neutrally stable manner. In addition, the zero eigenvalue may solely exist while the purely imaginary eigenvalues will naturally appear as a conjugate pair. In practice the neutrally stable solution will easily diverge as the numerical errors accumulate and the solution will drift even if restarted from a already converged steady state solution. 

Similar to the analysis for strong boundary conditions by Darmofal et al. \cite{darmofal2000eigenmode}, the coefficient matrix $\mathbf{C}_{2,int}$ contains the reflection properties of the adopted boundary conditions. However, here, it is the component matrices $\partial \hat{\mathbf{U}}_b^w/\partial \hat{\mathbf{U}}_{int}$ and $\partial \hat{\mathbf{U}}_b^e/\partial \hat{\mathbf{U}}_{int}$ rather than $\mathbf{C}_{2,int}$ should be checked to avoid entry cancellations. The zeros of $\left(\partial \hat{\mathbf{U}}_b^w/\partial \hat{\mathbf{U}}_{int}\right)_{13}$ and $\left(\partial \hat{\mathbf{U}}_b^w/\partial \hat{\mathbf{U}}_{int}\right)_{23}$ show that the inflow boundary condition has no reflection to the internal disturbance, while the zeros at $\left(\partial \hat{\mathbf{U}}_b^e/\partial \hat{\mathbf{U}}_{int}\right)_{31}$ and $\left(\partial \hat{\mathbf{U}}_b^e/\partial \hat{\mathbf{U}}_{int}\right)_{32}$ indicate a non-reflective outflow. Therefore, the combination of entropy-invariant compatible inflow and invariant compatible outflow is fully non-reflective in the normal direction.

\section{Stability analysis for pressure compatible inflows}
\label{sec::stabilityAnalysis}
We now analyzes the eigenvalues for $\mathbf{C}_{int}$ to determine a stable pressure compatible inflow for the DG solver. Three possible candidates are constructed and their stability performances analyzed. We only examine the coefficient matrix $\mathbf{C}_{int}$ since the opposite relation in Eq. (\ref{eq::LinStability1D_U_requirement_equivalent}) always hold.

\subsection{Entropy-pressure compatible inflow}
\label{sec::inflow_SP}
Neglecting the superscript $w$, the entropy-pressure compatible inflow at the west boundary in Fig. \ref{Fig_1D_Q} is given by 
\begin{equation}
\label{eq::SP_BC_1}
  p_{b} = p_{ref}, \hspace{0.5 cm}
  S_{b} = S_{ref}
\end{equation}
which directly implies 
\begin{equation}
\label{eq::SP_BC_2}
  \rho_{b} = \rho_{ref}, \hspace{0.5 cm}
  c_{b}    = c_{ref}.
\end{equation}

The third condition which needs to be specified according to the characteristic relations in Eq. (\ref{eq::characteristic_relation})
\begin{equation}
\label{eq::SP_BC_tmp_1}
  R_{b}^- = u_{b} - \frac{2}{\gamma-1} c_{b} =  R_{int}^-
\end{equation}
Then the boundary state velocity and right-propagating characteristic are computed as
\begin{equation}
\label{eq::SP_BC_tmp_2}
  u_{b} = R_{int}^- + \frac{2}{\gamma-1} c_{ref},
\end{equation}
\begin{equation}
\label{eq::SP_BC_tmp_3}
  R_{b}^+ = u_{b} + \frac{2}{\gamma-1} c_{b} = R_{int}^- + \frac{4}{\gamma-1} c_{ref} = R_{int}^- + \left(R_{ref}^+ - R_{ref}^-. \right)
\end{equation}
Therefore, the boundary state Riemann variables take the form
\begin{equation}
  \hat{\mathbf{U}}_{b} =\left(
    \begin{array}{c}
      R_{b}^+ \\  R_{b}^0 \\ R_{b}^-
    \end{array} \right) = \left(
    \begin{array}{c}
      R_{int}^- + R_{ref}^+ - R_{ref}^- \\  R_{ref}^0  \\ R_{int}^-
    \end{array} \right)
\end{equation}

As mentioned earlier, using the invariant compatible Riemann outflow, the second part of coefficient matrix $\mathbf{C}_{int}^{SP}$ is
\begin{equation}
\label{eq::SP_I_int}
  \mathbf{C}_{2,int}^{SP} = 
  \frac{\partial \hat{\mathbf{U}}_b^w}{\partial \hat{\mathbf{U}}_{int}} -  \frac{\partial \hat{\mathbf{U}}_b^e}{\partial \hat{\mathbf{U}}_{int}} = \left(
  \begin{array}{c c c}
    0 & 0 & 1 \\
    0 & 0 & 0 \\
    0 & 0 & 1
 \end{array} \right) - \left(
  \begin{array}{c c c}
    1 & 0 & 0 \\
    0 & 1 & 0 \\
    0 & 0 & 0
 \end{array} \right) = \left(
  \begin{array}{c c c}
    -1 &  0 & 1 \\
     0 & -1 & 0 \\
     0 &  0 & 1
 \end{array} \right)
\end{equation}
with which the eigenvalues for $\mathbf{C}_{int}^{SP}$ are
\begin{equation}
  \lambda_1^{SP} = -(c+u), \hspace{0.5 cm}
  \lambda_2^{SP} = -u, \hspace{0.5 cm}
  \lambda_3^{SP} = -(c-u)
\end{equation}
Since all of the eigenvalues are negative for subsonic flows, the combination of the entropy-pressure inflow and the invariable compatible outflow leads to a stable system.

It is worth noting that the entropy-pressure compatible inflow is able to enforce more conditions in multiple-dimensional cases, where the tangential velocity components are enforced by the reference state. The density compatibility in Eq. (\ref{eq::SP_BC_tmp_2}) makes the boundary state tangential momentum components also match the reference state.

\subsection{Velocity-pressure compatible inflow}
\label{sec::inflow_UP}
Following the same notation, the velocity-pressure compatible inflow satisfies
\begin{equation}
\label{eq::UP_BC_1}
  u_{b} = u_{ref}, \hspace{0.5 cm}
  p_{b} = p_{ref}
\end{equation}
Then the boundary state density is computed using the left-pointing characteristic as
\begin{equation}
\label{eq::UP_BC_2}
  \rho_{b} = \gamma \left(\frac{2}{\gamma-1} \right)^2 \frac{p_{ref}}{\left(u_{ref}-R_{int}^-\right)^2}
\end{equation}
Therefore, the boundary state characteristic variables are given by
\begin{equation}
\label{eq::UP_BC_3}
  R_{b}^+ = u_{b} + \frac{2}{\gamma-1} c_{b} = 2u_{ref} - R_{int}^- = \left(R_{ref}^+ + R_{ref}^- \right) - R_{int}^-
\end{equation}
\begin{equation}
  R_{b}^0 = \frac{p_{b}}{\rho_{b}^\gamma} = \left[ \frac{ \left( R_{ref}^+ + R_{ref}^- \right) -2 R_{int}^-}{ R_{ref}^+ - R_{ref}^-} \right]^{2\gamma} R_{ref}^0
\label{eq::UP_BC_4}
\end{equation}
which leads to
\begin{equation}
\label{eq::UP_I_int}
\begin{aligned}
  \mathbf{C}_{2,int}^{UP} = 
  \frac{\partial \hat{\mathbf{U}}_b^w}{\partial \hat{\mathbf{U}}_{int}} -  \frac{\partial \hat{\mathbf{U}}_b^e}{\partial \hat{\mathbf{U}}_{int}} & = \left(
  \begin{array}{c c c}
    0 & 0 & -1 \\
    0 & 0 & -(\gamma-1)\frac{c}{\rho^{\gamma-1}} \\
    0 & 0 & 1
 \end{array} \right) - \left(
  \begin{array}{c c c}
    1 & 0 & 0 \\
    0 & 1 & 0 \\
    0 & 0 & 0
 \end{array} \right) \\ & = \left(
  \begin{array}{c c c}
    -1 &  0 & -1 \\
     0 & -1 & -(\gamma-1)\frac{c}{\rho^{\gamma-1}} \\
     0 &  0 & 1
 \end{array} \right)
\end{aligned}
\end{equation}
and the eigenvalues of  $\mathbf{C}_{int}^{UP}$ are
\begin{equation}
  \lambda_1^{UP} = -(c+u), \hspace{0.5 cm}
  \lambda_{2,3}^{UP} = \pm \sqrt{-u(c-u)}
\end{equation}
where $\lambda_{2,3}^{UP}$ are purely imaginary for subsonic flows. The first order linear analysis, indicates that part of the internal disturbance will keep oscillating and therefore the velocity-pressure compatible inflow is not desirable for a steady state solution.

\subsection{Momentum-pressure compatible inflow}
\label{sec::inflow_MP}
Similarly, a momentum and velocity boundary state can be enforced by the conditions
\begin{equation}
\label{eq::MP_BC_1}
  \rho_{b} u_{b} = \rho_{ref} u_{ref}, \hspace{0.5 cm}
  p_{b} = p_{ref}. 
\end{equation}
By introducing a auxiliary variable, $\eta$, the boundary state density, velocity, and speed of sound take the form
\begin{equation}
\label{eq::MP_BC_2}
  \rho_{b} = \eta^{-2} \rho_{ref}, \hspace{0.5 cm}
  u_{b} = \eta^2 u_{ref}, \hspace{0.5 cm}
  c_{b} = \eta c_{ref}
\end{equation}
where
\begin{equation}
\label{eq::MP_BC_3}
  \eta = \sqrt{\frac{u_{b}}{u_{ref}}} = \frac{1}{\gamma-1} \frac{c_{ref}}{u_{ref}} - \sqrt{\left( \frac{1}{\gamma-1} \frac{c_{ref}}{u_{ref}} \right)^2 + \frac{R_{int}^-}{u_{ref}}  }.
\end{equation}
Therefore, the boundary state characteristic variables are
\begin{equation}
\label{eq::MP_BC_4}
  R_{b}^+ = u_{b} + \frac{2}{\gamma-1} c_{b} = \frac{\eta^2}{2} \left(R_{ref}^+ + R_{ref}^- \right) + \frac{\eta}{2}\left(R_{ref}^+ - R_{ref}^-\right),
\end{equation}
\begin{equation}
\label{eq::MP_BC_5}
  R_{b}^0 = \eta^{2\gamma} R_{ref}^0.
\end{equation}
The second part of coefficient matrix is therefore given by
\begin{equation}
\begin{aligned}
  \mathbf{C}_{2,int}^{MP} = 
  \frac{\partial \hat{\mathbf{U}}_b^w}{\partial \hat{\mathbf{U}}_{int}} -
  \frac{\partial \hat{\mathbf{U}}_b^e}{\partial \hat{\mathbf{U}}_{int}} & = \left(
  \begin{array}{c c c}
    0 & 0 & -2 \frac{c+(\gamma-1)u}{c-(\gamma-1)u} \\
    0 & 0 & -2 (\gamma-1) \frac{c^2}{\rho^{\gamma-1} \left[c-(\gamma-1)u\right]} \\
    0 & 0 & 1
 \end{array} \right) - \left(
  \begin{array}{c c c}
    1 & 0 & 0 \\
    0 & 1 & 0 \\
    0 & 0 & 0
 \end{array} \right) \\ & = \left(
  \begin{array}{c c c}
    -1 &  0 & -2 \frac{c+(\gamma-1)u}{c-(\gamma-1)u} \\
     0 & -1 & -2 (\gamma-1) \frac{c^2}{\rho^{\gamma-1} \left[c-(\gamma-1)u\right]} \\
     0 &  0 & 1
 \end{array} \right).
\label{M-p+stdOut_in}
\end{aligned}
\end{equation}
Subsequently the eigenvalues of $\mathbf{C}_{int}^{MP}$ are 
\begin{equation}
  \lambda_1^{MP} = -(c+u), \hspace{0.5 cm}
  \lambda_{2,3}^{MP} = \frac{\pm \sqrt{\sigma} - cu + c^2 + \gamma c u}{2(c+u-\gamma u)}
\end{equation}
where
\begin{equation}
  \sigma = c^4 + 2(\gamma-3) c^3 u + \gamma(\gamma+6) c^2 u^2 - 3c^2 u^2 + 4(1-\gamma^2) c u^3 + 4(\gamma -1)^2 u^4.
\end{equation}

\begin{figure}[htbp]
  \centering
  \includegraphics[width=10cm]{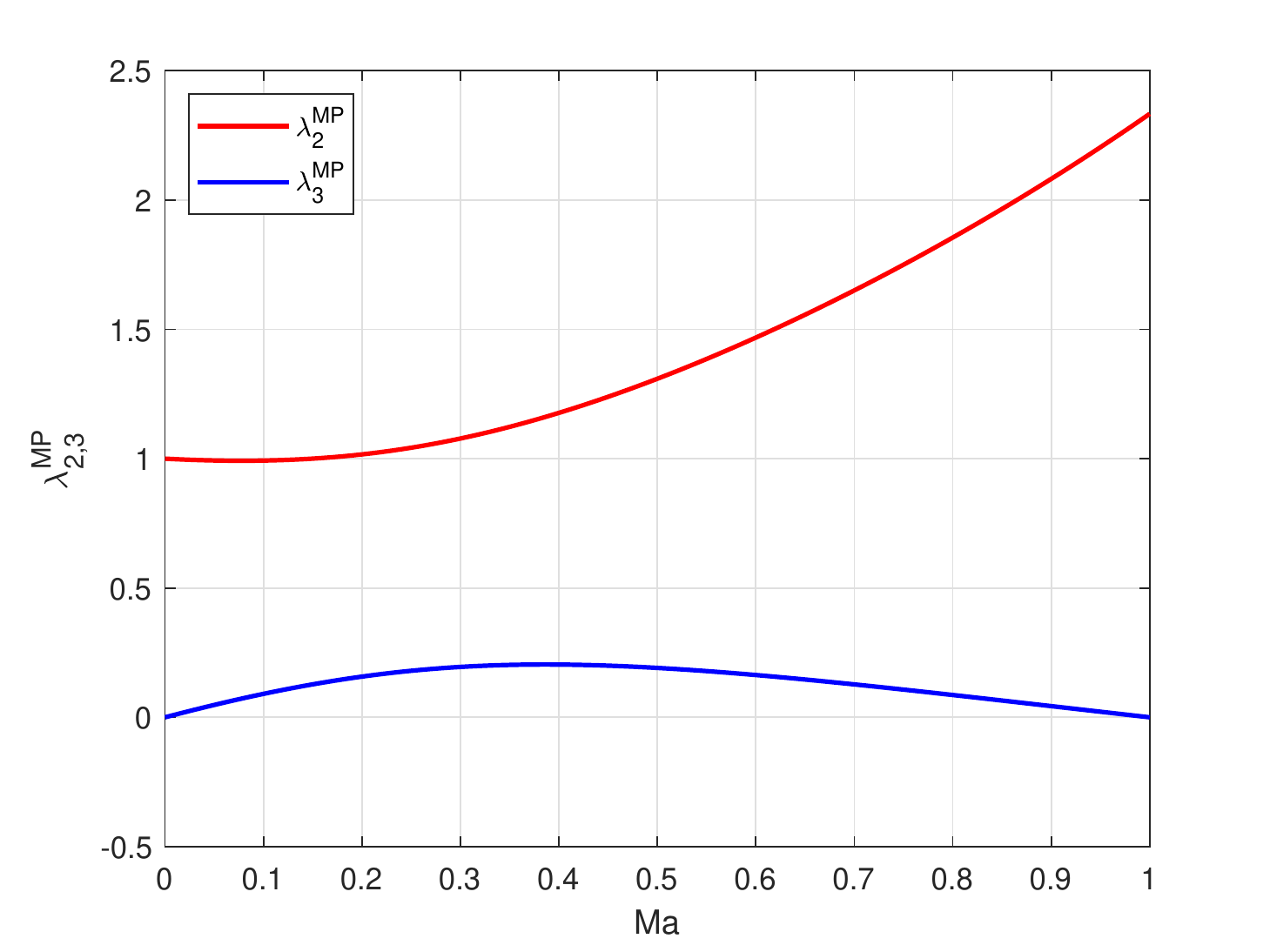}
  \caption{Variation of $\lambda_{2,3}^{MP}$ for subsonic flows.}
  \label{Fig_eigenvalues_MP}
\end{figure}

Since the second and third eigenvalues have complicated expressions, their values are examined numerically at different subsonic Mach numbers $Ma$, as shown in Fig.  \ref{Fig_eigenvalues_MP}. We observe that these two eigenvalues are always real and positive, leading to a unstable solution and so the adopted boundary conditions are unstable.

\section{Stability analysis of inflow entropy modification}
\label{sec::stabilityForInflowEntropyModification}

In Section \ref{sec::stabilityAnalysis} we demonstrated  that the only stable inflow boundary condition among the three candidates is the entropy-pressure compatible one. The similarity it shared with the entropy-invariant compatible inflow is that the boundary state entropy is kept the same as the reference state value. Using the similar method, entropy-velocity, density-velocity, and entropy-total enthalpy compatible inflows can be constructed and examined. The stability analysis result of these inflow boundary conditions are listed in Table. \ref{tab::stability_for_inflow_BCs}. Similarly, the entropy compatible conditions have negative eigenvalues for $\mathbf{C}_{int}$ while the others suffer stability issues from  non-negative eigenvalues. These observations imply the entropy is a essential quantity, and violating entropy compatibility is likely to cause problems. We therefore provide more 
detailed analysis for entropy modification at inflow.

\begin{table}[htbp]
\label{tab::stability_for_inflow_BCs}
  \centering
  \caption{Stability for different inflow boundary conditions}
    \begin{tabular}{c c c c c c}
    \toprule
        & Inflow compatibility & Outflow compatibility & Stability \\  
    \midrule
      1 & Entropy-invariant      & Invariant  & $\surd$  \\
      2 & Entropy-pressure       & Invariant  & $\surd$  \\
      3 & Velocity-pressure      & Invariant  & $\times$ \\
      4 & Momentum-pressure      & Invariant  & $\times$ \\
      5 & Entropy-velocity       & Invariant  & $\surd$  \\
      6 & Density-velocity       & Invariant  & $\times$ \\
      7 & Entropy-total enthalpy & Invariant  & $\surd$  \\
    \bottomrule
    \end{tabular}
\end{table}

We continue to follow the setting in Fig. \ref{Fig_1D_Q} where the west boundary is the inflow, and again the invariant compatible outflow at the east boundary is adopted. Since only the outward-propagating Riemann invariant may be used to construct the boundary state at the inflow, the second part coefficient matrix takes the form
\begin{equation}
\label{eq::S_modification_1}
  \mathbf{C}_{2,int} = 
  \frac{\partial \hat{\mathbf{U}}_b^w}{\partial \hat{\mathbf{U}}_{int}} -  \frac{\partial \hat{\mathbf{U}}_b^e}{\partial \hat{\mathbf{U}}_{int}} = \left(
  \begin{array}{c c c}
    0 & 0 & a \\
    0 & 0 & b \\
    0 & 0 & 1
 \end{array} \right) - \left(
  \begin{array}{c c c}
    1 & 0 & 0 \\
    0 & 1 & 0 \\
    0 & 0 & 0
 \end{array} \right) = \left(
  \begin{array}{c c c}
    -1 &  0 & a \\
     0 & -1 & b \\
     0 &  0 & 1
  \end{array} \right)
\end{equation}
where $a$ and $b$ are entries representing the contribution of the internal state. Then the characteristic polynomial for the coefficient matrix is
\begin{equation}
\label{eq::S_eigenvalue_1}
  \left[\lambda + (c+u)\right] \left\{ \lambda^2+\left[c+\frac{c^2}{\gamma(\gamma-1)S} b\right] \lambda +u(c-u) \right\}=0
\end{equation}
where $\lambda_1 = -(c+u)$ is always negative and $\lambda_{2,3}$ normalized by speed of sound are given by
\begin{equation}
\label{eq::S_eigenvalue_2}
  \bar{\lambda}_{2,3} = \frac{\lambda_{2,3}}{c} = \frac{-d \pm \sqrt{d^2-\beta}}{2}
\end{equation}
where $Ma=u/c$ is the Mach number, $\beta=4Ma \left( 1-Ma \right)$ is always a positive parameter for subsonic flows, $d$ is a normalized parameter
\begin{equation}
\label{eq::S_eigenvalue_3}
  d = 1+\frac{c}{\gamma(\gamma-1)S} b
\end{equation}

\begin{figure}[htbp]
  \centering
  \includegraphics[width=10cm]{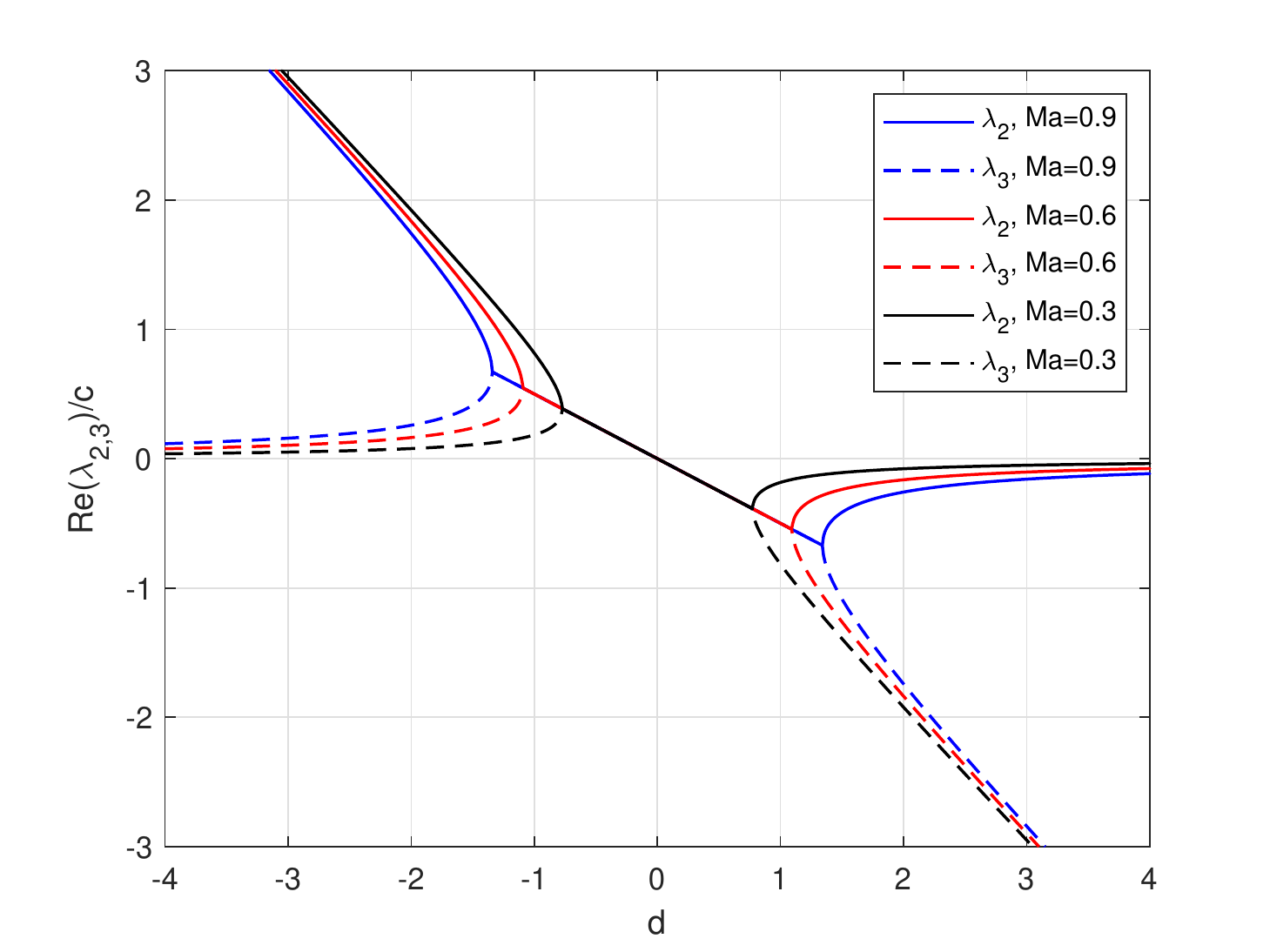}
  \caption{Normalized real part of $\lambda_{2,3}$ changing with parameter $d$ at different three Mach numbers: $Ma=0.3$ (black), $0.6$ (red), and $0.9$ (blue). The solid line represents $\lambda_2$ and the dashed line represents $\lambda_3$.}
  \label{Fig_lambda_vs_d}
\end{figure}

Eqs. (\ref{eq::S_eigenvalue_2}) and (\ref{eq::S_eigenvalue_3}) show that $\lambda_{2,3}$ are only related to the entry $b$ while the entry $a$ has nothing to do with stability. This allows the boundary state Riemann invariant $R^+_b$ to be modified as desired without introducing any singularity to the system. On the other hand, the modification of the boundary state entropy $R_b^0$ needs to follow some restrictions. From Eq. (\ref{eq::S_eigenvalue_2}) we can tell that the normalized real part of $\lambda_{2,3}$ is always negative for subsonic flows at $d>0$. This point is more clearly shown in Fig. \ref{Fig_lambda_vs_d} where $\lambda_{2,3}$ changes with $d$ at different Mach numbers. The right branches of the curves lead to negative real parts, corresponding to
\begin{equation}
\label{eq::S_eigenvalue_4}
  b > - \frac{\gamma(\gamma-1)S}{c}
\end{equation}
which needs to be satisfied for the boundary state entropy modification. Since the boundary state entropy equals to the ghost state entropy for an inflow boundary condition, Eq. (\ref{eq::S_eigenvalue_4}) is essentially the requirement to design the ghost state. 

There are two special cases. If $b=0$, the eigenvalues recover the ideal values as $\lambda_{2}=-u$ and $\lambda_3=-(c-u)$, which is the case for all the entropy compatible inflow. This indicates that as long as the entropy compatibility is satisfied at inflow, another inflow compatible quantity can  be chosen whilst maintaining a stable system. The other case is when $b = -\gamma(\gamma-1)S/c$, which is the case for the velocity-pressure compatible Riemann inflow, $\lambda_{2,3}$ are purely imaginary conjugates, leading to oscillating disturbances.

\section{Ill-posedness for entropy-pressure inflow at the presence of a stagnation point}
\label{sec::illPosedness_2D}

In the previous sections, the 1D eigenvalue analysis based on a steady and uniform baseflow shows that the entropy-pressure compatible inflow together with invariant compatible outflow leads to a stable system. However, multi-dimensional numerical tests highlight the emergence of a divergent solution for an inviscid simulations and strong oscillations appearing in viscous simulations. The divergence and oscillations are observed to initiate in 
the region of the stagnation point, for example a the leading edge of an aerofoil (see Section \ref{sec::results}). To understand this issue, we analyse a standard 2D squared domain ($\Omega = \left(x_1,x_2\right) \in [-1,1] \otimes [-1,1]$), where wall boundary condition are imposed on one of the boundaries. Due to the presence of the wall, the baseflow of the linearized system can no longer be assumed uniform, which is obviously a key difference compared with the previous 1D analysis. 
\begin{figure}[htbp]
  \centering
  \includegraphics[width=12cm]{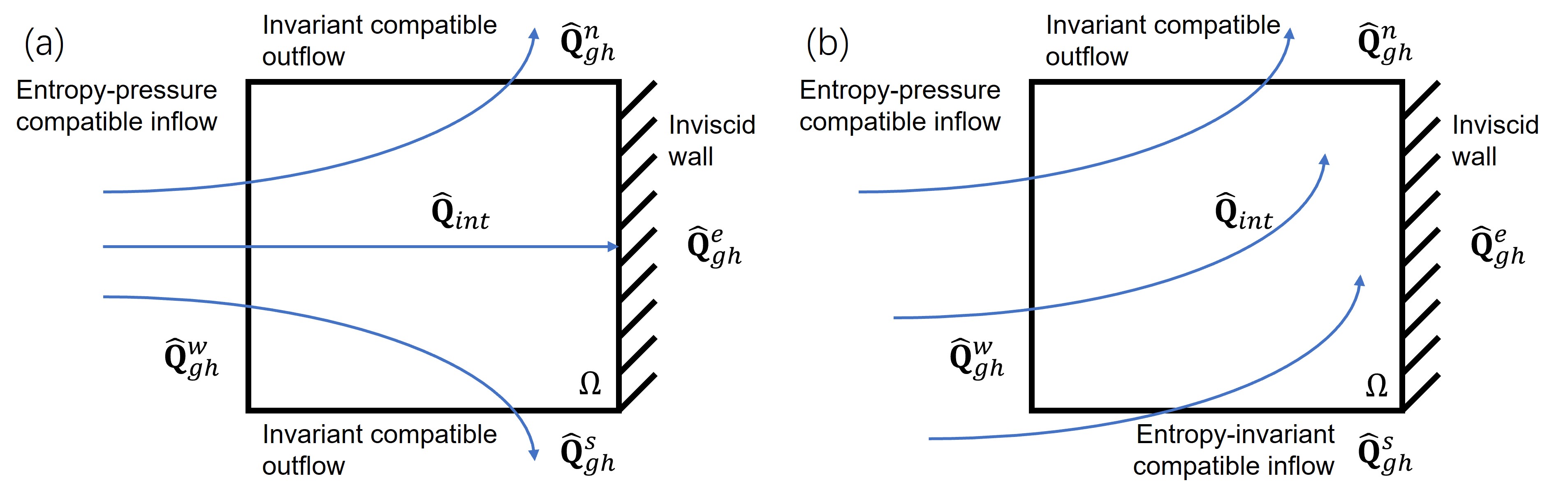} 
  \caption{Baseflow and boundary conditions for 2D stability analysis near to a wall: (a) with a stagnation point and (b) without a stagnation point.}
  \label{Fig_2D_Q_stagnation}
\end{figure}

Fig. \ref{Fig_2D_Q_stagnation} schematically shows a representative squared domain with associated boundary conditions in two cases with (a) and without (b) a stagnation point on the right wall. We next construct a 2D linearized system using the DG approximation of the 2D Euler equations, which take the conservation form
\begin{equation}
  \frac{\partial \mathbf{Q}}{\partial t} + 
  \frac{\partial \mathbf{F}_1(\mathbf{Q})}{\partial x_1} + 
  \frac{\partial \mathbf{F}_2(\mathbf{Q})}{\partial x_2} = \mathbf{0}
\end{equation}
where $\mathbf{Q}$ is the state vector of 2D conservative variables, $\mathbf{F}_1$ and $\mathbf{F}_2$ are the vectors of the inviscid flux
\begin{equation}
\label{eq::Euler2D}
  \mathbf{Q} =\left(
    \begin{array}{c}
      \rho \\ \rho u_1 \\ \rho u_2 \\ E
    \end{array} \right), \hspace{0.5 cm}
  \mathbf{F}_1 =\left(
    \begin{array}{c}
      \rho u_1 \\ \rho {u_1}^2 + p \\ \rho u_1 u_2 \\ u_1 (E + p)
    \end{array} \right), \hspace{0.5 cm}
  \mathbf{F}_2 =\left(
    \begin{array}{c}
      \rho u_2 \\ \rho u_1 u_2 \\ \rho {u_2}^2 + p \\u_2 (E + p)
    \end{array} \right). 
\end{equation}
In the above, $u_1$ and $u_2$ are the velocity components in the $x_1$ and $x_2$-directions, respectively, and $E = \frac{1}{\gamma - 1} p + \frac{1}{2} \rho({u_1}^2 + {u_2}^2)$. Others symbols are the same as in Eq. (\ref{eq::Euler1D_differenial}).

Following a similar analysis as performed in Section \ref{sec::Linearized_DG_for_Euler1D} and \ref{sec::appendix_A}, the domain $\Omega$ is approximated by a single quadrilateral DG element as 
\begin{equation}
\label{eq::Euler2D_DG_1}
  \int_{\Omega} \phi \frac{\partial \mathbf{Q}}{\partial t} d\Omega +
  \int_{\Gamma} \phi \left(\tilde{\mathbf{f}} \cdot \mathbf{n} \right) d\Gamma -
  \int_{\Omega} \nabla \phi \cdot \mathbf{F} d\Omega = \mathbf{0}
\end{equation}
where $\Gamma =\partial \Omega$ is the boundary of the domain, $\mathbf{F} = \left(\mathbf{F}_1,\mathbf{F}_2 \right)^T$ is the block vector for inviscid flux \cite{hindenlang2020stability} while $\tilde{\mathbf{f}} = \left(\tilde{\mathbf{f}}_1,\tilde{\mathbf{f}}_2 \right)^T$ is the corresponding block vector of numerical inviscid fluxes. 


The expanded form of Eq. (\ref{eq::Euler2D_DG_1}) on the squared domain in Fig. \ref{Fig_2D_Q_stagnation}(a) is 
\begin{equation}
\label{eq::Euler2D_DG_2}
  \int_{\Omega} \phi \frac{\partial \mathbf{Q}}{\partial t} d\Omega =
  \int_{\Gamma^w} \phi \tilde{\mathbf{f}}_1^w  d x_2 -
  \int_{\Gamma^e} \phi \tilde{\mathbf{f}}_1^e  d x_2 +
  \int_{\Gamma^s} \phi \tilde{\mathbf{f}}_2^s  d x_1 -
  \int_{\Gamma^n} \phi \tilde{\mathbf{f}}_2^n  d x_1 +
  \int_{\Omega} \nabla \phi \cdot \mathbf{F} d\Omega
\end{equation}
where the subscript $n$ and $s$ denote north and south, respectively. Due to the presence of the wall the baseflow is no longer uniform so that a two-dimensional approximation of the following form is considered
\begin{equation}
\label{eq::fullApproximation}
  \mathbf{Q} \simeq \sum_{p=0}^{P_1} \sum_{q=0}^{P_2} \hat{\mathbf{Q}}_{pq}(t) \phi_{pq}(x_1,x_2), \hspace{0.5 cm}
  \phi_{pq}(x_1,x_2) = \phi_{p}(x_1) \phi_{q}(x_2)
\end{equation} 
where $P_1$ and $P_2$ are expansion order in the orthogonal directions, and  typically are set to be equal to each other $P_1=P_2=P$. In the following analysis we choose to adopt a Lagrange nodal basis function for both $\phi_p$ and $\phi_q$ of the form
\begin{equation}
\label{eq::basisFunction_nodal}
  \phi_p(x) = \left\{ 
  \begin{aligned} 
    & 1, \hspace{3.7 cm} \quad x=x_p,  \\
    & \frac{(x-1)(x+1)L'_P(x)}{P(P+1)L_P(x_p)(x_p-x)}, \quad \text{otherwise},\\
  \end{aligned} \right. \quad 0 \le p \le P
\end{equation}
where $x_p$ represents the Gauss-Lobatto-Legendre points including both end-points of the interval. 
For clarity, we use the semi-discrete form by only numerically integrating the boundary terms
\begin{equation}
\begin{aligned}
\label{eq::Euler2D_DG_3}
  \int_{\Omega} \phi \frac{\partial \mathbf{Q}_{int}}{\partial t} d\Omega & = 
  \sum_{i=0}^{P} \phi w_i \tilde{\mathbf{f}}_{1,i}^w 
  \left( \hat{\mathbf{Q}}_{ref,i}^w,\hat{\mathbf{Q}}_{int,i}^w \right)  -
  \sum_{i=0}^{P} \phi w_i \tilde{\mathbf{f}}_{1,i}^e 
  \left( \hat{\mathbf{Q}}_{ref,i}^e,\hat{\mathbf{Q}}_{int,i}^e \right) \\ & +
  \sum_{i=0}^{P} \phi w_i \tilde{\mathbf{f}}_{2,i}^s 
  \left( \hat{\mathbf{Q}}_{ref,i}^s,\hat{\mathbf{Q}}_{int,i}^s \right)  -
  \sum_{i=0}^{P} \phi w_i \tilde{\mathbf{f}}_{2,i}^n 
  \left( \hat{\mathbf{Q}}_{ref,i}^n,\hat{\mathbf{Q}}_{int,i}^n \right) \\ & +
  \int_{\Omega} \nabla \phi \cdot \mathbf{F} d\Omega 
\end{aligned}
\end{equation}
where the numerical flux going through the wall, $\tilde{\mathbf{f}}_{1,i}^e$, is constructed using the internal state and no-penetration condition provided through the reference state $\hat{\mathbf{Q}}_{ref,i}^e$. Since the no-penetration condition, or $\hat{\mathbf{Q}}_{ref,i}^e$, is unrelated with the outer simulation, it is not disturbed and therefore in the linearized system the disturbance on the reference state for the east boundary is zero. Moreover, the last term in Eq. (\ref{eq::Euler2D_DG_3}) is volumic flux integral where only the internal state is involved. The above leads to that the contribution by the reference state disturbances have two terms fewer than the contribution by the internal state disturbances.

To construct the 2D linearized system with normalized and neat coefficients, the following variables are introduced
\begin{equation}
\label{eq::quasiCharacteristicVars}
  \mathbf{V} =\left(
    \begin{array}{c c c c}
      \frac{2}{\gamma-1} c, &  u_1, & u_2, & S
    \end{array} \right)^T
\end{equation}
which are referred to as quasi-characteristic variables since their linear combinations give the 1D characteristic variables in Eq. (\ref{eq::characteristicVars}). The subsequent linear system for Fig. \ref{Fig_2D_Q_stagnation}(a) takes the form (see \ref{sec::appendix_B} for details)
\begin{equation}
\label{eq::Linearized_DG_Euler2D_V}
  \int_{\Omega} \phi \frac{\partial \mathbf{Q}}{\partial \mathbf{V}}
  \frac{\partial \left(\delta \mathbf{V}_{int} \right)}{\partial t} d\Omega = RHS_{int}\left(\hat{\mathbf{V}}, \delta \hat{\mathbf{V}}_{int} \right) + RHS_{ref}\left(\hat{\mathbf{V}}, \delta \hat{\mathbf{V}}_{ref} \right)
\end{equation}
where the contributions by internal state disturbances and reference state disturbances are respectively included in the right-hand side terms $RHS_{int}$ and $RHS_{ref}$, are given by
\begin{equation}
\label{eq::Linearized_DG_Euler2D_V_int}
\begin{aligned}
  & RHS_{int}\left(\hat{\mathbf{V}}, \delta \hat{\mathbf{V}}_{int} \right) = \\
  & \sum_{i=0}^{P} \phi w_i \left[ 
  \frac{\partial \tilde{\mathbf{f}}_{1,i}^w}{\partial \hat{\mathbf{Q}}_{b,i}^w} \frac{\partial \hat{\mathbf{Q}}_{b,i}^w}{\partial \hat{\mathbf{V}}_{b,i}^w} \frac{\partial \hat{\mathbf{V}}_{b,i}^w}{\partial \hat{\mathbf{V}}_{int,i}^w}
  \delta \hat{\mathbf{V}}_{int,i}^w \right] 
  - \sum_{i=0}^{P} \phi w_i \left[ 
  \frac{\partial \tilde{\mathbf{f}}_{1,i}^e}{\partial \hat{\mathbf{Q}}_{b,i}^e} \frac{\partial \hat{\mathbf{Q}}_{b,i}^e}{\partial \hat{\mathbf{V}}_{b,i}^e} 
  \frac{\partial \hat{\mathbf{V}}_{b,i}^e}{\partial \hat{\mathbf{V}}_{int,i}^e} \delta \hat{\mathbf{V}}_{int,i}^e \right] \\ 
  + & \sum_{i=0}^{P} \phi w_i \left[ 
  \frac{\partial \tilde{\mathbf{f}}_{1,i}^s}{\partial \hat{\mathbf{Q}}_{b,i}^s} \frac{\partial \hat{\mathbf{Q}}_{b,i}^s}{\partial \hat{\mathbf{V}}_{b,i}^s} \frac{\partial \hat{\mathbf{V}}_{b,i}^s}{\partial \hat{\mathbf{V}}_{int,i}^s} \delta \hat{\mathbf{V}}_{int,i}^s \right] 
  - \sum_{i=0}^{P} \phi w_i \left[ 
  \frac{\partial \tilde{\mathbf{f}}_{1,i}^n}{\partial \hat{\mathbf{Q}}_{b,i}^n} 
  \frac{\partial \hat{\mathbf{Q}}_{b,i}^n}{\partial \hat{\mathbf{V}}_{b,i}^n} \frac{\partial \hat{\mathbf{V}}_{b,i}^n}{\partial \hat{\mathbf{V}}_{int,i}^n} \delta \hat{\mathbf{V}}_{int,i}^n \right] \\
  + & \int_{\Omega} \nabla \phi \cdot \frac{\partial \mathbf{F}}{\partial \mathbf{Q}} \frac{\partial \mathbf{Q}}{\partial \mathbf{V}} \delta \mathbf{V}_{int} d\Omega
\end{aligned}
\end{equation}
and
\begin{equation}
\label{eq::Linearized_DG_Euler2D_V_ref}
\begin{aligned}
  & RHS_{ref}\left(\hat{\mathbf{V}}, \delta \hat{\mathbf{V}}_{ref} \right) = \\
    & \sum_{i=0}^{P} \phi w_i \left[ 
  \frac{\partial \tilde{\mathbf{f}}_{1,i}^w}{\partial \hat{\mathbf{Q}}_{b,i}^w} \frac{\partial \hat{\mathbf{Q}}_{b,i}^w}{\partial \hat{\mathbf{V}}_{b,i}^w} \frac{\partial \hat{\mathbf{V}}_{b,i}^w}{\partial \hat{\mathbf{V}}_{ref,i}^w}
  \delta \hat{\mathbf{V}}_{ref,i}^w \right] 
  + \sum_{i=0}^{P} \phi w_i \left[ 
  \frac{\partial \tilde{\mathbf{f}}_{1,i}^s}{\partial \hat{\mathbf{Q}}_{b,i}^s} \frac{\partial \hat{\mathbf{Q}}_{b,i}^s}{\partial \hat{\mathbf{V}}_{b,i}^s} \frac{\partial \hat{\mathbf{V}}_{b,i}^s}{\partial \hat{\mathbf{V}}_{ref,i}^s} \delta \hat{\mathbf{V}}_{ref,i}^s \right]  \\
  - & \sum_{i=0}^{P} \phi w_i \left[ 
  \frac{\partial \tilde{\mathbf{f}}_{1,i}^n}{\partial \hat{\mathbf{Q}}_{b,i}^n} 
  \frac{\partial \hat{\mathbf{Q}}_{b,i}^n}{\partial \hat{\mathbf{V}}_{b,i}^n} \frac{\partial \hat{\mathbf{V}}_{b,i}^n}{\partial \hat{\mathbf{V}}_{ref,i}^n} \delta \hat{\mathbf{V}}_{ref,i}^n \right]
\end{aligned}
\end{equation}
In the above, due to the presence of the wall, neither the baseflow nor the disturbances have a uniform distribution. Therefore the right-hand side terms cannot be further simplified to form two coefficient matrices as we have derived in the 1D analysis. So from the conventional point of view, the eigenvalue analysis for the coefficient matrix only with respect to the internal disturbance cannot be applied to this system. However, the instability issue can be shown by checking the responses to the reference state disturbances through the  derivative matrix, i.e. $\partial \hat{\mathbf{V}}_{b,i}^l / \partial \hat{\mathbf{V}}_{ref,i}^l$ ($l=w,s,n$) which arises in  each term of $RHS_{ref}$. Considering the boundary conditions adopted, the boundary states for the east, north, and south boundaries are given by
\begin{equation}
  \hat{\mathbf{V}}_b^w =\left(
  \begin{array}{c}
    \frac{2}{\gamma-1} c_{ref}^w \\  u_{1,int}^w - \frac{2}{\gamma-1} c_{int}^w + \frac{2}{\gamma-1} c_{ref}^w \\  u_{2,ref}^w \\ S_{ref}^w
  \end{array} \right)
\end{equation}
and
\begin{equation}
  \hat{\mathbf{V}}_b^{n/s} = \left(
  \begin{array}{c}
    \frac{1}{2} \left[ \left( u_{2,int}^{n/s}+ \frac{2}{\gamma-1} c_{int}^{n/s} \right) - \left( u_{2,ref}^{n/s} - \frac{2}{\gamma-1}c_{ref}^{n/s} \right) \right] \\ 
    u_{1,int}^{n/s} \\
    \frac{1}{2} \left[ \left( u_{2,int}^{n/s}+ \frac{2}{\gamma-1}c_{int}^{n/s} \right) + \left( u_{2,ref}^{n/s} - \frac{2}{\gamma-1}c_{ref}^{n/s}\right) \right]  \\ 
    S_{int}^{n/s}
  \end{array} \right)
\end{equation}
where the subscript "n/s" denotes either north or south boundary, and the velocity tangential to the boundary is enforced to be the upstream value according to the feature of split Riemann problem \cite{Toro2009}.  The respective derivative matrices of boundary state with respect to the internal and reference states are
\begin{equation}
  \frac{\partial \hat{\mathbf{V}}_b^w}{\partial \hat{\mathbf{V}}_{int}^w} = \left(
  \begin{array}{c c c c}
     0 & 0 & 0 & 0 \\
    -1 & 1 & 0 & 0 \\
     0 & 0 & 0 & 0 \\
     0 & 0 & 0 & 0
 \end{array} \right), \hspace{0.5 cm}
 \frac{\partial \hat{\mathbf{V}}_b^w}{\partial \hat{\mathbf{V}}_{ref}^w} = \left(
  \begin{array}{c : c : c c}
     1 & 0 & 0 & 0 \\
     1 & 0 & 0 & 0 \\
     0 & 0 & 1 & 0 \\
     0 & 0 & 0 & 1
 \end{array} \right)
\label{eq::SP_2D}
\end{equation}
\begin{equation}
  \frac{\partial \hat{\mathbf{V}}_b^{n/s}}{\partial \hat{\mathbf{V}}_{int}^{n/s}} = \left(
  \begin{array}{c c c c}
     \frac{1}{2} & 0 & \frac{1}{2} & 0 \\
     0 & 1 & 0 & 0 \\
     \frac{1}{2} & 0 & \frac{1}{2} & 0 \\
     0 & 0 & 0 & 1
 \end{array} \right), \hspace{0.5 cm}
 \frac{\partial \hat{\mathbf{V}}_b^{n/s}}{\partial \hat{\mathbf{V}}_{ref}^{n/s}} = \left(
  \begin{array}{c : c : c c}
      \frac{1}{2} & 0 & -\frac{1}{2} & 0 \\
     0 & 0 & 0 & 0 \\
     -\frac{1}{2} & 0 & \frac{1}{2} & 0 \\
     0 & 0 & 0 & 0
 \end{array} \right)
\label{eq::SI_2D}
\end{equation}
We notice that the entries in the second column in $\partial \hat{\mathbf{V}}_{b}^w / \partial \hat{\mathbf{V}}_{ref}^w$ and $\partial \hat{\mathbf{V}}_{b}^{n/s} / \partial \hat{\mathbf{V}}_{ref}^{n/s}$ are all zero. Since the disturbances of $u_1$ on the reference states are introduced into the domain through these entries, the zeros infer that the linearized system will not be influenced by the disturbance of $u_1$ from any of the boundaries. This non-physical condition leaves an under-determined $u_1$-field, and therefore the boundary condition we have adopted are ill-posed for this problem. 

Although the influence of the wall and stagnation point can hardly be analyzed in a 1D case (otherwise there is no inflow for the steady baseflow), this issue can still be understood in a 1D perspective. For simplicity, the velocity, speed of sound, and entropy are chosen as the independent variables. For the weakly imposed wall boundary condition, the ghost state velocity is specified as the opposite value of the internal state velocity while the ghost state density and energy are set equal to the internal state counterparts \cite{mengaldo2014guide}. In such a manner, the boundary state velocity is zero and the speed of sound equals to the internal state value, which makes the inward-propagating Riemann invariant equal to the internal state speed of sound times $2/(\gamma-1)$. At the inflow boundary, when the entropy-invariant compatible inflow is adopted, the entropy and the linear combination of velocity and speed of sound are specified. This boundary conditions combination leads to a solvable problem for each field. However, when adopting the entropy-pressure compatible inflow, the entropy and pressure provide the speed of sound at the inflow boundary but leaves a free velocity, which is not influenced by the speed of sound reflected back from the wall.

The ill-posedness issue in Fig. \ref{Fig_2D_Q_stagnation}(a) can be addressed by introducing a boundary condition where the $u_1$ velocity component can be imposed on the solution. This leads to the scenario in Fig. \ref{Fig_2D_Q_stagnation}(b), where the south boundary is set as an inflow and therefore there is no stagnation point on the wall. If the entropy-invariant inflow is present for the south boundary, the derivative matrix takes the form
\begin{equation}
  \frac{\partial \hat{\mathbf{V}}_b^s}{\partial \hat{\mathbf{V}}_{ref}^s} = \left(
  \begin{array}{c : c : c c}
     \frac{1}{2} & 0 & \frac{1}{2} & 0 \\
     0 & 1 & 0 & 0 \\
     \frac{1}{2} & 0 & \frac{1}{2} & 0 \\
     0 & 0 & 0 & 1
  \end{array} \right)
\label{eq::SI_in_2D_2}
\end{equation}
In this case the non-zero entry in the second column enables the disturbance on the reference state $u_1$-velocity component go into the domain to determine the field and therefore the problem becomes well-posed.

\section{Results}
\label{sec::results}

Having identified out the stable entropy-pressure compatible inflow and understood its potential singularity issue around a stagnation point, we next provide an example for pressure compatible simulation in the reduced domain on the CRM-NLF model \cite{CRM-NLF}. The Navier-Stokes equations are solved using the open-source spectral/hp framework  \emph{Nektar++} \cite{cantwell2015nektar++,moxey2020nektar++}. 

\begin{figure}[htbp]
  \centering
  \includegraphics[width=8cm]{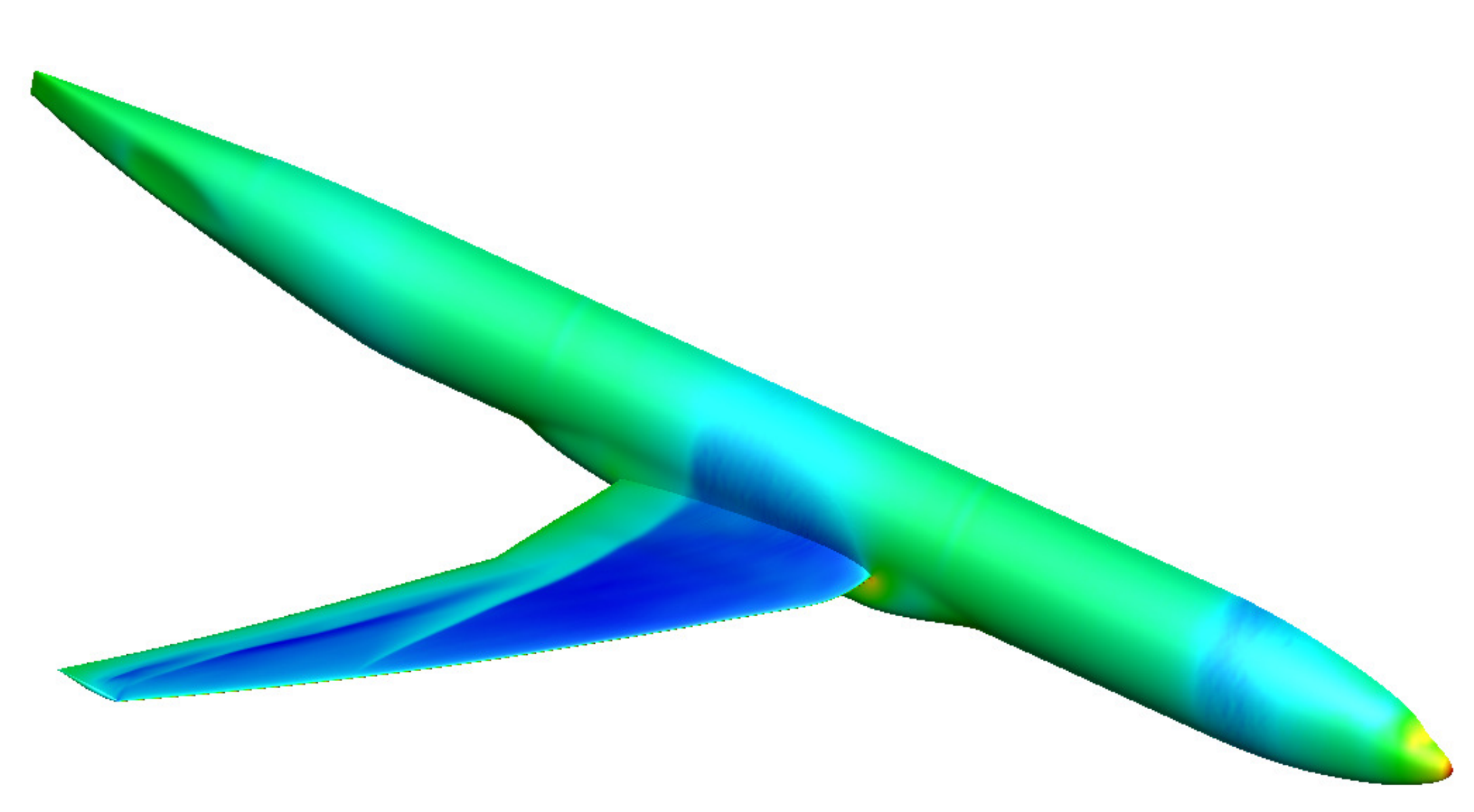}
  \caption{Pressure distribution in the RANS simulation over full the CRM-NLF model.}
  \label{Fig_CRM_RANS}
\end{figure}

\begin{figure}[htbp]
  \centering
  \includegraphics[width=10cm]{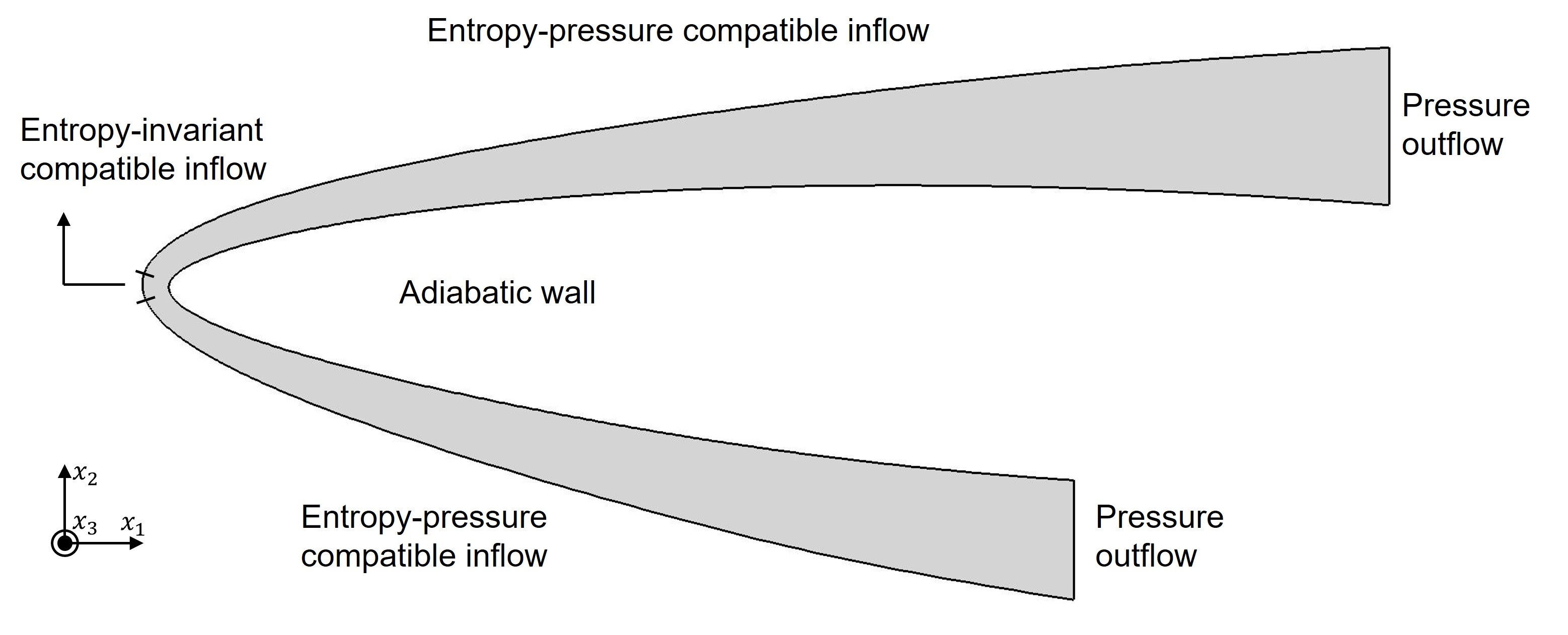}
  \caption{Boundary conditions for the reduced domain in the CRM-NLF simulation. The periodic boundary condition is adopted in the $x_3$-direction for the 3D simulation.}
  \label{Fig_airfoil_BCs}
\end{figure}

To generate the background fields, a transonic RANS simulation over the full wing-fuselage geometry is first carried out under  the conditions provided in Table. \ref{Tab_freestream_conditions}, where $\alpha$ is the angle of attack, $L$ is the reference length. Fig. \ref{Fig_CRM_RANS} shows the pressure distribution on the surface where two shocks on the wing can be observed. We therefore set the outflow boundary of the reduced domain at the upstream of the shock. The reduced domain is obtained on the slice in the wall-normal direction passing the leading-edge of section D of the CRM-NLF model \cite{CRM-NLF}. Fig. \ref{Fig_airfoil_BCs} shows the reduced domain together with the boundary condition strategy for advection terms. To avoid the stagnation-induced ill-posedness issue discussed in Section \ref{sec::illPosedness_2D}, the entropy-invariant compatible inflow is applied at the nose region of the domain, covering the streamline that passes the stagnation point. The pressure outflow (see Ref. \cite{mengaldo2014guide}) rather than the invariant compatible outflow is adopted since the pressure interpolated from the outer RANS simulation is considered a more reliable quantity inside the boundary layer. We compare the results using the entropy-pressure compatible inflow and results using the standard Riemann-invariant inflow boundary conditions, where a single inflow boundary condition is applied over the full inflow boundary. To carry out a quasi-3D simulation, the periodic boundary condition is enforced in the $x_3$-direction. The boundary condition treatments for viscous fluxes are the same as those in Ref. \cite{mengaldo2014guide}.

\begin{table}[htbp]
  \centering
  \caption{Freestream conditions for the RANS simulation of the CRM-NLF model.}
  \label{Tab_freestream_conditions}
    \begin{tabular}{c c c c c c c c}
    \toprule
    $Ma$ & $Re_L$ & $\alpha$ [deg] & $T_{\infty}$ [K] & $U_{\infty}$ [m/s] & $\rho_{\infty}$ [kg/m$^3$] & $L$ [m]\\  
    \midrule
    $0.856$ & $8.5\times10^6$ & $1.5$ & $277.1$ & $285.7$ & 3.343 & 0.154678\\
    \bottomrule
    \end{tabular}
\end{table}

\begin{figure}[htbp]
  \centering
  \subfigure[Outer RANS simulation]{
    \label{fig_pressure_RANS}
    \includegraphics[width=5.85cm]{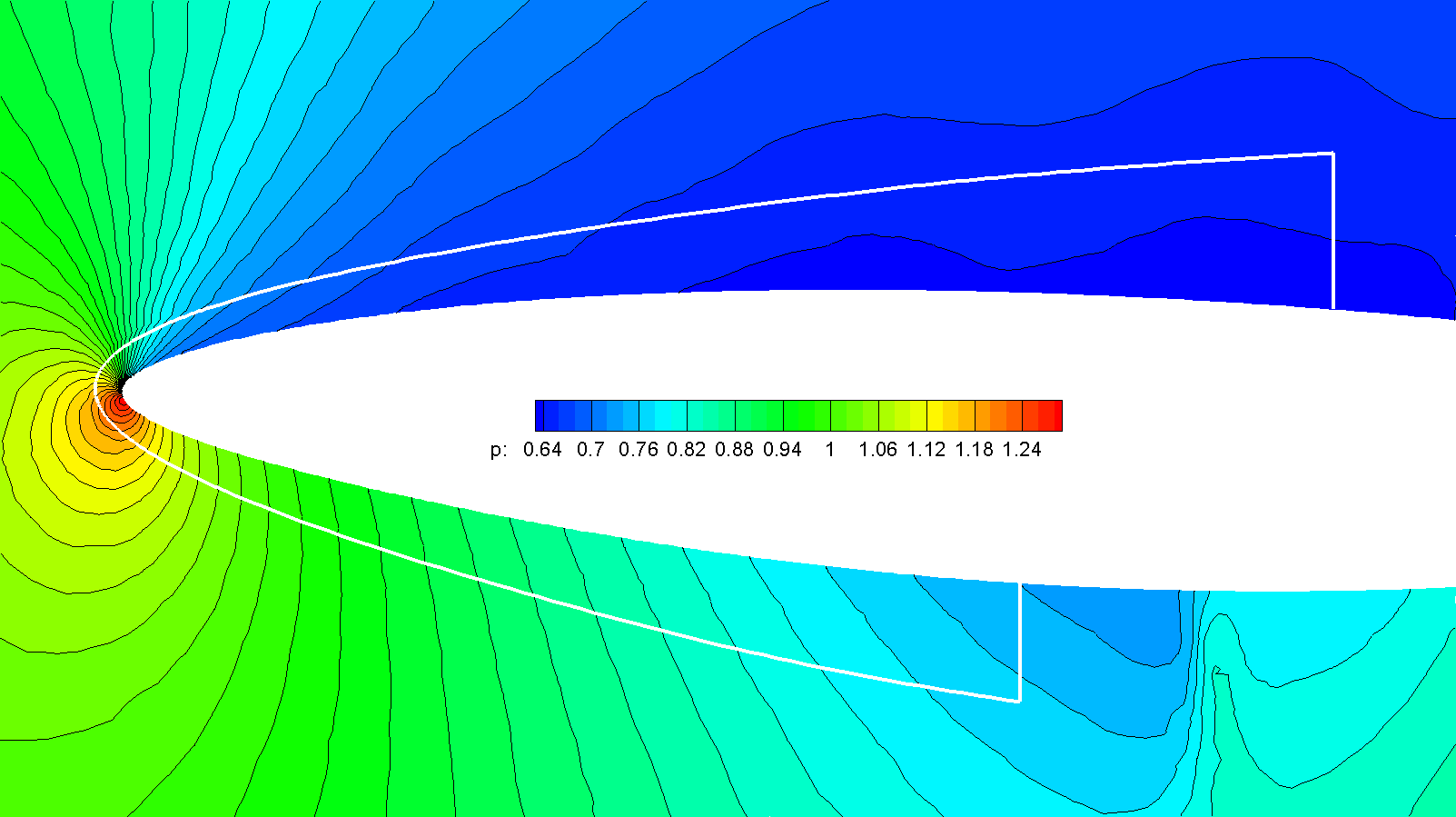}}
  \subfigure[Entropy-pressure compatible inflow]{
    \label{fig_pressure:b}
    \includegraphics[width=5.85cm]{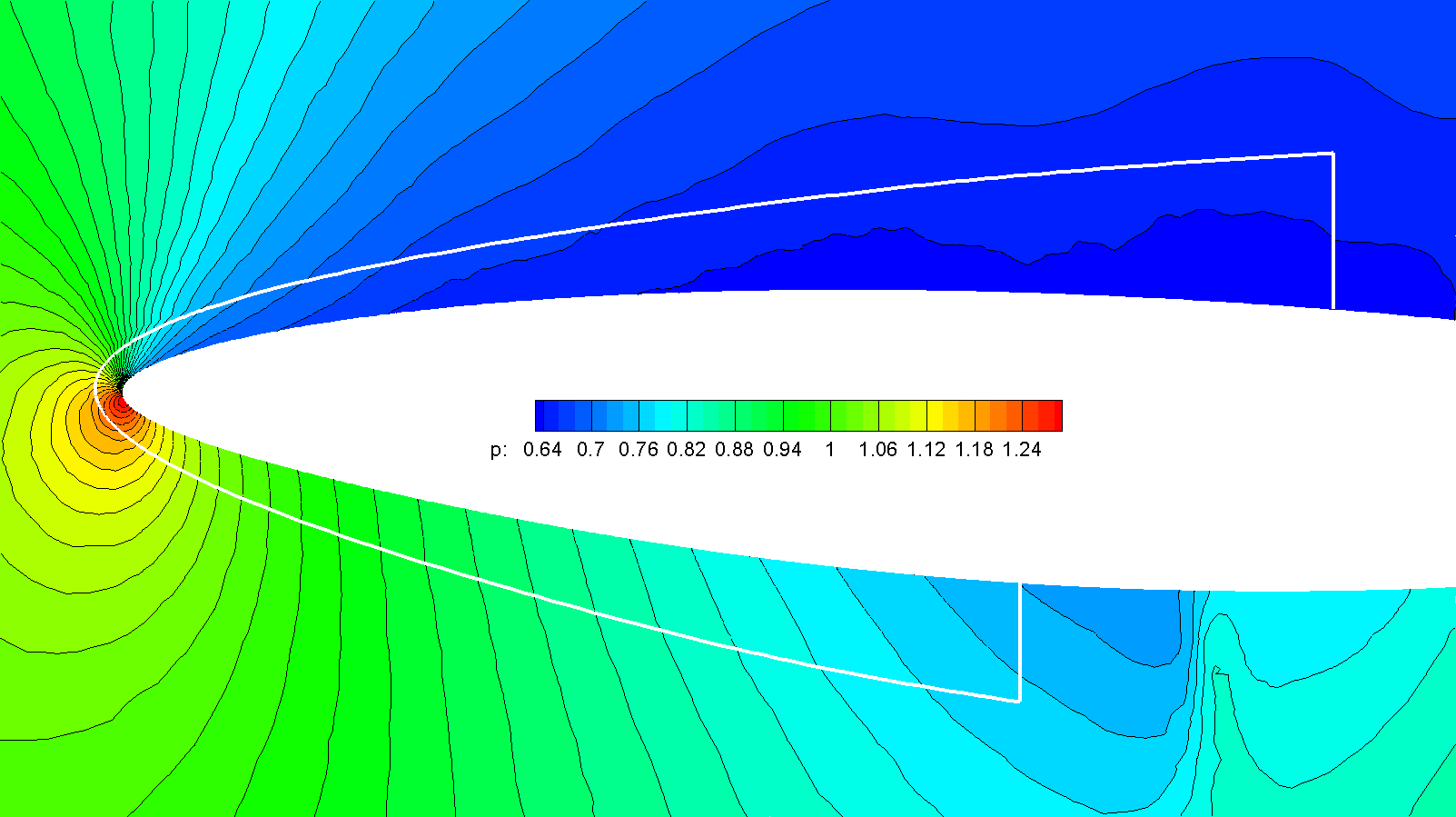}}
  \subfigure[Entropy-invariant compatible inflow]{
    \label{fig_pressure:c}
    \includegraphics[width=5.85cm]{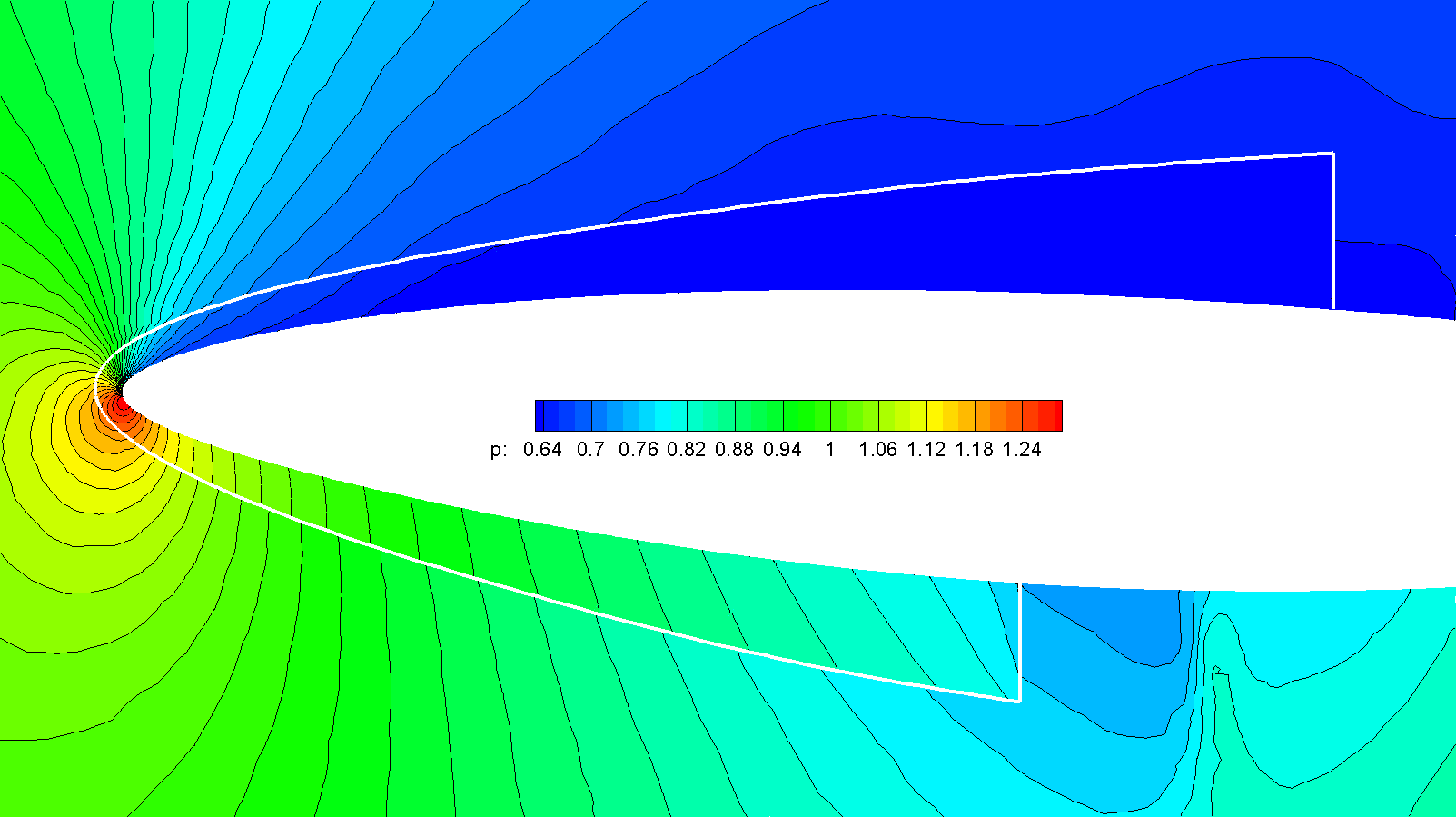}}
  \subfigure[Entropy-enthalpy compatible inflow]{
    \label{fig_pressure:d}
    \includegraphics[width=5.85cm]{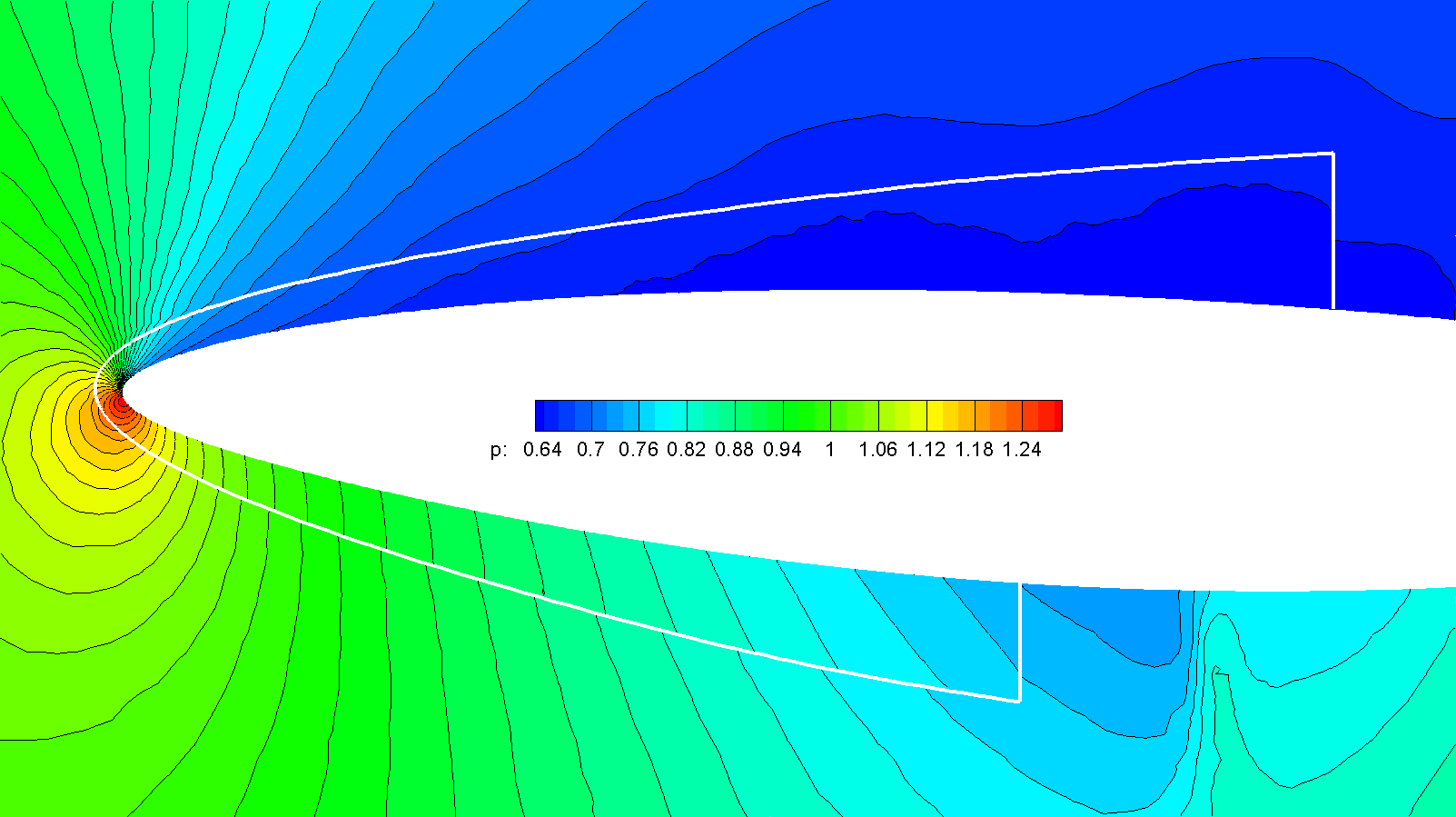}}
  \caption{Pressure contours comparison for using different boundary conditions.}
\label{Fig_pressure_contour}
\end{figure}

In Fig. \ref{Fig_pressure_contour} we compare the pressure distributions when using different inflow boundary conditions. The result for entropy-pressure compatible inflow is given in Fig. \ref{fig_pressure:b} where the contour lines perfectly match the pressure contours of the RANS field. Fig. \ref{fig_pressure:c} and Fig. \ref{fig_pressure:d} provide the pressure distributions without pressure compatibility enforced. Significant mismatching in contour lines is observed in the result for the entropy-invariant compatible inflow in Fig. \ref{fig_pressure:c}, where the invariant compatible outflow rather than pressure outflow needs to be adopted for a stable simulation because of the large pressure incompatibility. Fig. \ref{fig_pressure:d} shows the pressure distribution for using the entropy-enthalpy compatible inflow. This result is better than that for entropy-invariant compatible inflow but some contour lines still unable to match the RANS data.

\begin{figure}[htbp]
  \centering
  \subfigure[Overall]{
    \label{fig_Cp:a}
    \includegraphics[width=10cm]{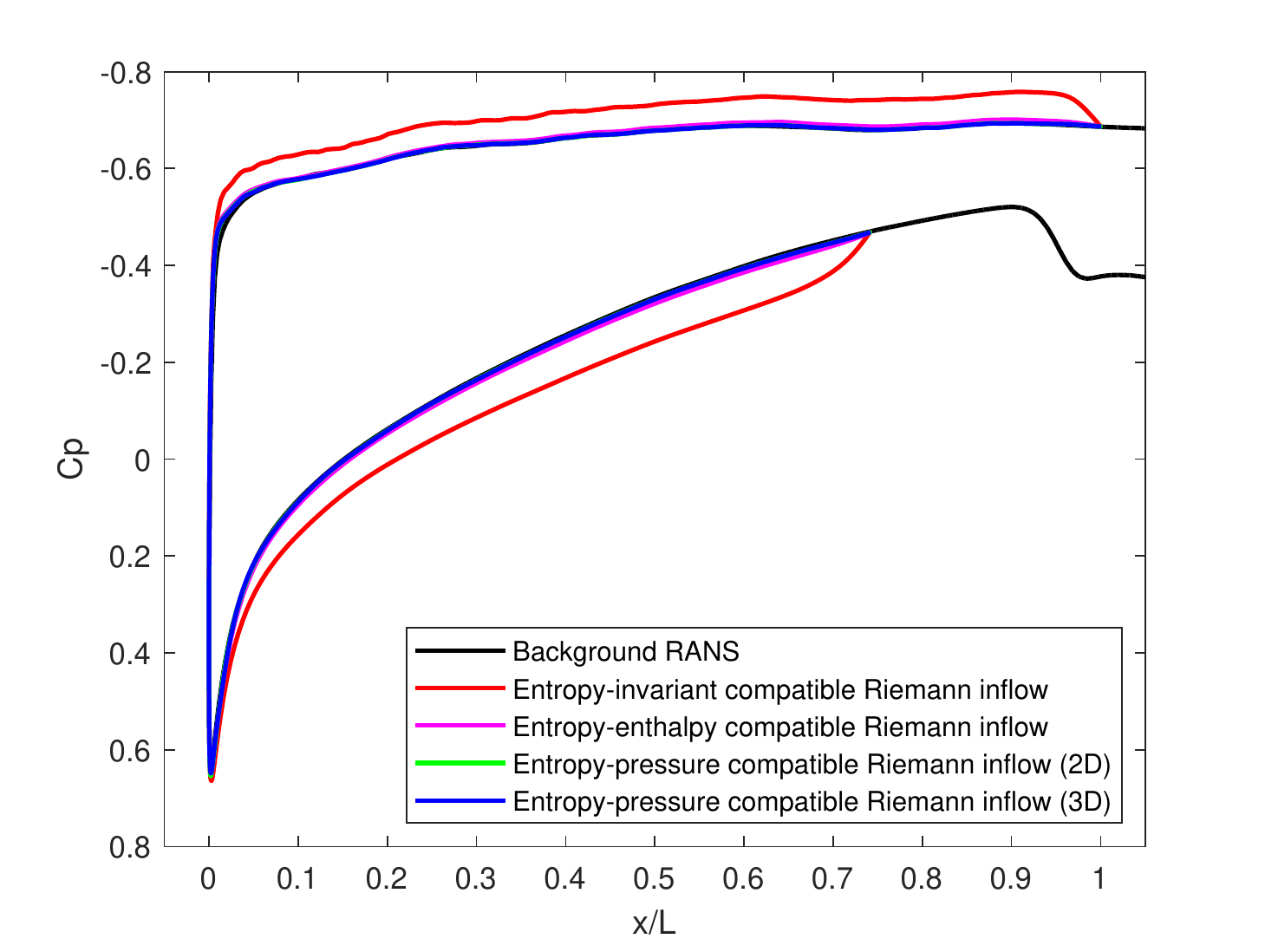}}
  \subfigure[Zoomed in]{
    \label{fig_Cp:b}
    \includegraphics[width=10cm]{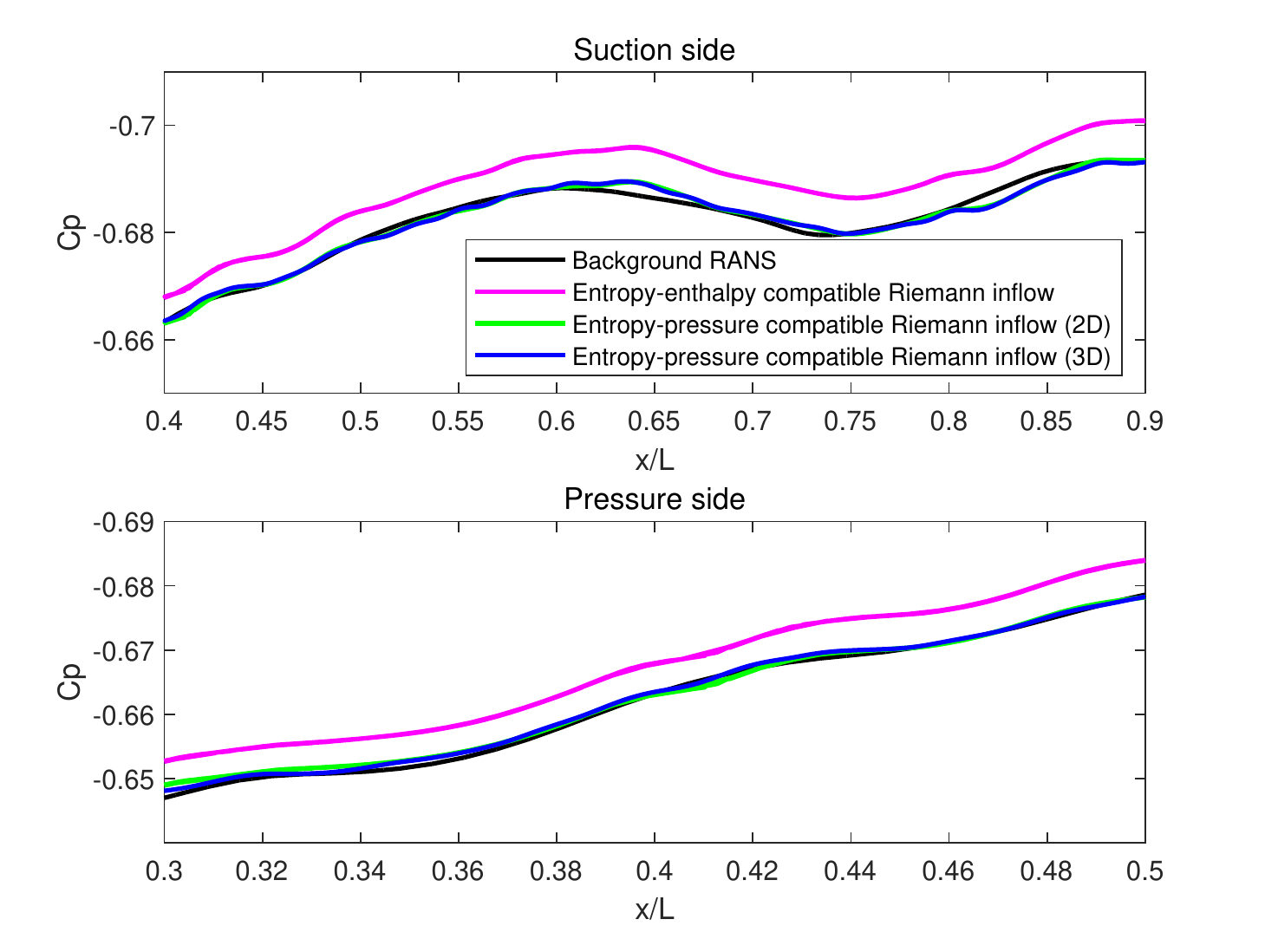}}
  \caption{Comparison for Cp distributions.}
\label{Fig_Cp}
\end{figure}

Fig. \ref{Fig_Cp} provides more quantitative comparisons for pressure coefficient (Cp) distributions on both upper and lower surfaces of the wing section. In Fig. \ref{fig_Cp:a} the discrepancy in the curve by entropy-invariant compatible inflow is readdressed, and the results by entropy-enthalpy compatible inflow and entropy-pressure compatible inflow are closer to each other. However, the zoomed-in comparison in Fig. \ref{fig_Cp:b} shows superiority of the entropy-pressure compatible inflow. To have a more well-rounded comparison, a quasi-3D simulation is also carried out using the entropy-pressure compatible inflow and the Cp distribution is provided. Excellent agreements with the RANS data are observed for both 2D and quasi-3D results in the comparison, indicating the effectiveness of the entropy-pressure compatible inflow.


\section{Conclusion}
\label{sec::conclusion}

To carry out high fidelity simulation in a reduced domain which is embedded in a outer domain where a lower fidelity model such as RANS is applied, appropriate boundary condition enforcement is required. This is particularly the case when the high fidelity simulation is compressible but subsonic, since  not all the data interpolated from the outer simulation can be enforced at the inflow boundary of the reduced domain. Since the pressure load is typically well predicted by the low fidelity model and does not  vary much over the boundary layer, the pressure is considered a reliable quantity to maintain in the reduced domain. We have therefore determined the stable enforcement to achieve pressure compatibility at subsonic inflow boundary for DG compressible flow simulations.

Initially we derived the linearized DG approximation of 1D Euler equations using piecewise constant assmptions. The linearized system is considered in a single element since in DG simulations the boundary conditions are enforced in an individual element in form of numerical flux and the inflow boundary condition is enforced at the most upstream boundaries of the inflow boundary elements. To complete the analysis the invariant compatible outflow were considered as it is implicit the outflow condition for internal boundaries between the elements (e.g. the very upstream element and the adjacent downstream element) when the numerical flux is the solution from an exact Riemann solver.

The construction of numerical flux depends on the internal state (inside the element) and the reference state (which is usually the desired state) and so the  linearized system can be derived from these two perspectives by introducing disturbance on either the internal state or the reference state, generating two different matrix problems. Introducing disturbance on the internal state is a standard approach, which finally leads to a linear dynamic system. Similarly from the reference state disturbance point of view, the derived coefficient matrix essentially represents how the reference state disturbance are received by the internal state. In the 1D analysis, these two perspectives have a complementary relation. Therefore the stability  of the boundary conditions can be obtained by computing the eigenvalues of either of the matrices. For the coefficient matrix derived from the perturbed internal state, a stable boundary condition requires the three eigenvalues of the linearised matrix to  have negative real parts. A representative example is the entropy-invariant compatible inflow, which leads to the eigenvalues of well-known $-(c+u)$, $-u$, and $-(c-u)$. A special case is when some eigenvalues are zero, indicating the coefficient matrix is rank-deficit. This usually implies that the boundary conditions is no well-posed, and therefore are inappropriate. 

To achieve pressure compatibility at the inflow, three types inflow conditions were constructed and examined, namely entropy-pressure, velocity-pressure and momentum pressure compatible condiitons. It can be shown that only the entropy-pressure compatible inflow is stable while velocity-pressure compatible inflow is neutrally stable and the momentum-pressure compatible inflow is unstable. It is further possible to prove that the entropy is an important quantity to maintain in order to be compatible with the outer simulation,  However, as long as the entropy compatibility is maintained, any other quantity can be chosen to be enforced at the inflow. 

With the understanding for 1D stability analysis, the stable entropy-pressure compatible inflow is applied to the 2D simulations. However, the simulations were observed to easily diverge, which initiates around the region of a stagnation point. To understand this instability, a similar method was 
derived for the 2D Euler equations in a squared domain. Due to the presence of the stagnation point, the baseflow is no longer uniform and the distribution is unknown. The assumption that the disturbances introduced to different boundaries are equal is also dropped. The significant consequence is that boundary integration term cannot be combined to form a single matrix for eigenvalues analysis. From the perspective of disturbance introduced on to the internal state, the instability issue cannot easily be further analyzed. However, the other perspective where the disturbance is introduced on to the reference state reveals that a rank deficient matrix problem related to the boundary interior coupling in the final linearized system. This therefore demonstrate that the instability is arising from the ill-posedness of the adopted boundary conditions.

The solution to this numerical instability issue is to locally replace the entropy-pressure compatible inflow by some other stable inflow, which enforces the normal velocity either explicitly or implicitly, in the boundary region related to the streamline related to the stagnation point. In this work we use the entropy-invariant compatible inflow to avoid this issue. We perform a  numerical test based on the CRM-NLF model. A RANS simulation over the full wing-fuselage geometry is performed, and a high fidelity simulation in the reduced domain on the slice passing the leading-edge was generated. With the updated boundary conditions strategy, the simulation is stable and the Cp distribution agrees well with the RANS data.

\section*{Acknowledgments}
The authors acknowledge support from the Beijing Aircraft Technology Research Institute of COMAC from 2019 to 2021.

\bibliography{mybibfile}

\appendix
\section{Linearized DG approximation for 1D Euler equations}
\label{sec::appendix_A}

The integration of 1D Euler equation gives
\begin{equation}
\label{eq::app_A_Baseflow_1D}
  \frac{d \hat{\mathbf{Q}}}{d t} \Delta x = \tilde{\mathbf{f}}^{w} - \tilde{\mathbf{f}}^{e} 
\end{equation}
where $\tilde{\mathbf{f}}^w$ and $\tilde{\mathbf{f}}^e$ are numeriacl fluxes at west and east boundaries. Each of them is constructed by the internal state $\hat{\mathbf{Q}}_{int}$ and a reference state $\hat{\mathbf{Q}}_{ref}$. The linearized system can be derived by introducing perturbations to all of these states, leading to
\begin{equation}
\label{eq::app_A_Baseflow_perturbation_1D}
\begin{aligned}
  \frac{d \left( \hat{\mathbf{Q}} + \delta \hat{\mathbf{Q}}_{int} \right)}{d t} \Delta x 
  & = \left( \tilde{\mathbf{f}}^w + \frac{\partial \tilde{\mathbf{f}}^w}{\partial \hat{\mathbf{Q}}_{int}}  \delta \hat{\mathbf{Q}}_{int} + \frac{\partial \tilde{\mathbf{f}}^w}{\partial \hat{\mathbf{Q}}_{ref}^w}  \delta \hat{\mathbf{Q}}_{ref}^w \right) \\ 
  & - \left( \tilde{\mathbf{f}}^e  + \frac{\partial \tilde{\mathbf{f}}^e}{\partial \hat{\mathbf{Q}}_{int}} \delta \hat{\mathbf{Q}}_{int} + \frac{\partial \tilde{\mathbf{f}}^e}{\partial \hat{\mathbf{Q}}_{ref}^e} \delta \hat{\mathbf{Q}}_{ref}^e \right)
\end{aligned}
\end{equation}

By substituting Eq. (\ref{eq::app_A_Baseflow_1D}) from Eq. (\ref{eq::app_A_Baseflow_perturbation_1D}), and assuming the perturbation on both reference states are equal, $\delta \hat{\mathbf{Q}}_{ref}^w = \delta \hat{\mathbf{Q}}_{ref}^e=\hat{\mathbf{Q}}_{ref}$, the 1D linearized systems is given by
\begin{equation}
\label{eq::app_A_LinStability1D_Q}
  \frac{d \left(\delta \hat{\mathbf{Q}}_{int} \right)}{d t} \Delta x =
  \left( \frac{\partial \tilde{\mathbf{f}}^w}{\partial \hat{\mathbf{Q}}_{int}} - \frac{\partial \tilde{\mathbf{f}}^e}{\partial \hat{\mathbf{Q}}_{int}} \right)  \delta \hat{\mathbf{Q}}_{int} + \left( \frac{\partial \tilde{\mathbf{f}}^w}{\partial \hat{\mathbf{Q}}_{ref}^w}  - 
  \frac{\partial \tilde{\mathbf{f}}^e}{\partial \hat{\mathbf{Q}}_{ref}^e} \right) \delta \hat{\mathbf{Q}}_{ref}
\end{equation}
which can then be transformed into characteristic variables as
\begin{equation}
\label{eq::app_A_LinStability1D_U}
\begin{aligned}
  \frac{d }{d t} \left(\frac{\partial \hat{\mathbf{Q}}_{int}}{\partial \hat{\mathbf{U}}_{int}} \delta \hat{\mathbf{U}}_{int} \right) \Delta x & =   \left( 
  \frac{\partial \tilde{\mathbf{f}}^w}{\partial \hat{\mathbf{Q}}_b^w} \frac{\partial \hat{\mathbf{Q}}_b^w}{\partial \hat{\mathbf{U}}_b^w} \frac{\partial \hat{\mathbf{U}}_b^w}{\partial \hat{\mathbf{Q}}_{int}} - \frac{\partial \tilde{\mathbf{f}}^e}{\partial \hat{\mathbf{Q}}_b^e} \frac{\partial \hat{\mathbf{Q}}_b^e}{\partial \hat{\mathbf{U}}_b^e} \frac{\partial \hat{\mathbf{U}}_b^e}{\partial \hat{\mathbf{Q}}_{int}} 
  \right)
  \frac{\partial \hat{\mathbf{Q}}_{int}}{\partial \hat{\mathbf{U}}_{int}} 
  \delta \hat{\mathbf{U}}_{int} \\ & + \left( 
  \frac{\partial \tilde{\mathbf{f}}^w}{\partial \hat{\mathbf{Q}}_b^w} \frac{\partial \hat{\mathbf{Q}}_b^w}{\partial \hat{\mathbf{U}}_b^w} \frac{\partial \hat{\mathbf{U}}_b^w}{\partial \hat{\mathbf{Q}}_{ref}^w} - \frac{\partial \tilde{\mathbf{f}}^e}{\partial \hat{\mathbf{Q}}_b^e} \frac{\partial \hat{\mathbf{Q}}_b^e}{\partial \hat{\mathbf{U}}_b^e} \frac{\partial \hat{\mathbf{U}}_b^e}{\partial \hat{\mathbf{Q}}_{ref}^e} \right)  
  \frac{\partial \hat{\mathbf{Q}}_{ref}^e}{\partial \hat{\mathbf{U}}_{ref}^e} \delta \hat{\mathbf{U}}_{ref}
\end{aligned}
\end{equation}

Due to the assumption that the baseflow is constant, Eq. (\ref{eq::app_A_LinStability1D_U}) is simplified as
\begin{equation}
\label{eq::app_A_LinStability1D_U_2}
\begin{aligned}
  \frac{d }{d t} \left(\delta \hat{\mathbf{U}}_{int} \right) & =
  \frac{1}{\Delta x}
  \left(\frac{\partial \hat{\mathbf{Q}}}{\partial \hat{\mathbf{U}}} \right)^{-1}
  \frac{\partial \tilde{\mathbf{f}}}{\partial \hat{\mathbf{Q}}}
  \frac{\partial \hat{\mathbf{Q}}}{\partial \hat{\mathbf{U}}}
  \left(
  \frac{\partial \hat{\mathbf{U}}_b^w}{\partial \hat{\mathbf{U}}_{int}} -  \frac{\partial \hat{\mathbf{U}}_b^e}{\partial \hat{\mathbf{U}}_{int}} \right)  \delta \hat{\mathbf{U}}_{int} \\ & + 
  \frac{1}{\Delta x}
  \left(\frac{\partial \hat{\mathbf{Q}}}{\partial \hat{\mathbf{U}}} \right)^{-1}
  \frac{\partial \tilde{\mathbf{f}}}{\partial \hat{\mathbf{Q}}}
  \frac{\partial \hat{\mathbf{Q}}}{\partial \hat{\mathbf{U}}}
  \left( 
  \frac{\partial \hat{\mathbf{U}}_b^w}{\partial \hat{\mathbf{U}}_{ref}^w} -  \frac{\partial \hat{\mathbf{U}}_b^e}{\partial \hat{\mathbf{U}}_{ref}^e} \right)  \delta \hat{\mathbf{U}}_{ref}
\end{aligned}
\end{equation}
or
\begin{equation}
\label{eq::app_A_LinStability1D_U_3}
  \frac{d }{d t} \left( \delta \hat{\mathbf{U}}_{int} \right) =
  \frac{1}{\Delta x} \mathbf{C}_{int} \delta \hat{\mathbf{U}}_{int} + \frac{1}{\Delta x} \mathbf{C}_{ref} \delta \hat{\mathbf{U}}_{ref} 
\end{equation}
where the state vectors and matrices without subscript or superscript denote the baseflow quantities which are uniform for the 1D analysis.

\section{Linearized DG approximation for 2D Euler equations}
\label{sec::appendix_B}
The DG approximation for 2D Euler equation at the presence of wall on the east boundary (Fig.\ref{Fig_2D_Q_stagnation}) is derived from the integration form of 2D Euler equations
\begin{equation}
\label{eq::app_B_Euler2D_DG_baseflow}
  \int_{\Omega} \phi \frac{\partial \mathbf{Q}}{\partial t} d\Omega =
  \int_{\Gamma^w} \phi \tilde{\mathbf{f}}_1^w  d x_2 -
  \int_{\Gamma^e} \phi \tilde{\mathbf{f}}_1^e  d x_2 +
  \int_{\Gamma^s} \phi \tilde{\mathbf{f}}_2^s  d x_1 -
  \int_{\Gamma^n} \phi \tilde{\mathbf{f}}_2^n  d x_1 +
  \int_{\Omega} \nabla \phi \cdot \mathbf{F} d\Omega
\end{equation}

We use the semi-discrete form by only numerically integrating the boundary terms, the DG approximation reads
\begin{equation}
\begin{aligned}
\label{eq::app_B_Euler2D_DG_semiDiscrete}
  \int_{\Omega} \phi \frac{\partial \mathbf{Q}_{int}}{\partial t} d\Omega & = 
  \sum_{i=0}^{P} \phi w_i \tilde{\mathbf{f}}_{1,i}^w 
  \left( \hat{\mathbf{Q}}_{ref,i}^w,\hat{\mathbf{Q}}_{int,i}^w \right)  -
  \sum_{i=0}^{P} \phi w_i \tilde{\mathbf{f}}_{1,i}^e 
  \left( \hat{\mathbf{Q}}_{ref,i}^e,\hat{\mathbf{Q}}_{int,i}^e \right) \\ & +
  \sum_{i=0}^{P} \phi w_i \tilde{\mathbf{f}}_{2,i}^s 
  \left( \hat{\mathbf{Q}}_{ref,i}^s,\hat{\mathbf{Q}}_{int,i}^s \right)  -
  \sum_{i=0}^{P} \phi w_i \tilde{\mathbf{f}}_{2,i}^n 
  \left( \hat{\mathbf{Q}}_{ref,i}^n,\hat{\mathbf{Q}}_{int,i}^n \right) \\ & +
  \int_{\Omega} \nabla \phi \cdot \mathbf{F} d\Omega 
\end{aligned}
\end{equation}
where in the reference state for the east boundary (wall), $\hat{\mathbf{Q}}_{ref,i}^e$, the momentum in the normal direction is zero for no-penetration condition. Following the same procedures in \ref{sec::appendix_A}, the linearized system for introducing disturbances on all the states leads to 
\begin{equation}
\label{eq::app_B_Euler2D_DG_linearized}
\begin{aligned}
  & \int_{\Omega} \phi 
  \frac{\partial \left(\delta \mathbf{Q}_{int} \right)}{\partial t} d\Omega = \\
    & \sum_{i=0}^{P} \phi w_i \left[ 
  \frac{\partial \tilde{\mathbf{f}}_{1,i}^w}{\partial \hat{\mathbf{Q}}_{int,i}^w} \delta \hat{\mathbf{Q}}_{int,i}^w + 
  \frac{\partial \tilde{\mathbf{f}}_{1,i}^w}{\partial \hat{\mathbf{Q}}_{ref,i}^w} \delta \hat{\mathbf{Q}}_{ref,i}^w \right] \\
  - & \sum_{i=0}^{P} \phi w_i \left[ 
  \frac{\partial \tilde{\mathbf{f}}_{1,i}^e}{\partial \hat{\mathbf{Q}}_{int,i}^e} \delta \hat{\mathbf{Q}}_{int,i}^e + 
  \frac{\partial \tilde{\mathbf{f}}_{1,i}^e}{\partial \hat{\mathbf{Q}}_{ref,i}^e} \delta \hat{\mathbf{Q}}_{ref,i}^e \right] \\
  + & \sum_{i=0}^{P} \phi w_i \left[ 
  \frac{\partial \tilde{\mathbf{f}}_{1,i}^s}{\partial \hat{\mathbf{Q}}_{int,i}^s} \delta \hat{\mathbf{Q}}_{int,i}^s + 
  \frac{\partial \tilde{\mathbf{f}}_{1,i}^s}{\partial \hat{\mathbf{Q}}_{ref,i}^s} \delta \hat{\mathbf{Q}}_{ref,i}^s \right] \\
  - & \sum_{i=0}^{P} \phi w_i \left[ 
  \frac{\partial \tilde{\mathbf{f}}_{1,i}^n}{\partial \hat{\mathbf{Q}}_{int,i}^n} \delta \hat{\mathbf{Q}}_{int,i}^n + 
  \frac{\partial \tilde{\mathbf{f}}_{1,i}^n}{\partial \hat{\mathbf{Q}}_{ref,i}^n} \delta \hat{\mathbf{Q}}_{ref,i}^n \right] +
  \int_{\Omega} \nabla \phi \cdot \frac{\partial \mathbf{F}}{\partial \mathbf{Q}} \delta \mathbf{Q}_{int} d\Omega
\end{aligned}
\end{equation}
Since the $\hat{\mathbf{Q}}_{ref,i}^e$ will not be influenced by the outer simulation, we have $\delta \hat{\mathbf{Q}}_{ref,i}^e=\mathbf{0}$. The consequent linearized system for Fig. \ref{Fig_2D_Q_stagnation}(a) is derived as
\begin{equation}
\label{eq::app_B_Euler2D_DG_linearized_Q}
  \int_{\Omega} \phi 
  \frac{\partial \left(\delta \mathbf{Q}_{int} \right)}{\partial t} d\Omega = RHS_{int}\left(\hat{\mathbf{Q}}, \delta \hat{\mathbf{Q}}_{int} \right) + RHS_{ref}\left(\hat{\mathbf{Q}}, \delta \hat{\mathbf{Q}}_{ref} \right)
\end{equation}
which is then transformed into quasi-characteristic variables (Eq. (\ref{eq::quasiCharacteristicVars})) as
\begin{equation}
\label{eq::app_B_Euler2D_DG_linearized_V}
  \int_{\Omega} \phi \frac{\partial \mathbf{Q}}{\partial \mathbf{V}}
  \frac{\partial \left(\delta \mathbf{V}_{int} \right)}{\partial t} d\Omega = RHS_{int}\left(\hat{\mathbf{V}}, \delta \hat{\mathbf{V}}_{int} \right) + RHS_{ref}\left(\hat{\mathbf{V}}, \delta \hat{\mathbf{V}}_{ref} \right)
\end{equation}
where the right-hand side terms take the form
\begin{equation}
\label{eq::app_B_Euler2D_DG_linearized_int_V}
\begin{aligned}
  & RHS_{int}\left(\hat{\mathbf{V}}, \delta \hat{\mathbf{V}}_{int} \right) = \\
  & \sum_{i=0}^{P} \phi w_i \left[ 
  \frac{\partial \tilde{\mathbf{f}}_{1,i}^w}{\partial \hat{\mathbf{Q}}_{b,i}^w} \frac{\partial \hat{\mathbf{Q}}_{b,i}^w}{\partial \hat{\mathbf{V}}_{b,i}^w} \frac{\partial \hat{\mathbf{V}}_{b,i}^w}{\partial \hat{\mathbf{V}}_{int,i}^w}
  \delta \hat{\mathbf{V}}_{int,i}^w \right] 
  - \sum_{i=0}^{P} \phi w_i \left[ 
  \frac{\partial \tilde{\mathbf{f}}_{1,i}^e}{\partial \hat{\mathbf{Q}}_{b,i}^e} \frac{\partial \hat{\mathbf{Q}}_{b,i}^e}{\partial \hat{\mathbf{V}}_{b,i}^e} 
  \frac{\partial \hat{\mathbf{V}}_{b,i}^e}{\partial \hat{\mathbf{V}}_{int,i}^e} \delta \hat{\mathbf{V}}_{int,i}^e \right] \\ 
  + & \sum_{i=0}^{P} \phi w_i \left[ 
  \frac{\partial \tilde{\mathbf{f}}_{1,i}^s}{\partial \hat{\mathbf{Q}}_{b,i}^s} \frac{\partial \hat{\mathbf{Q}}_{b,i}^s}{\partial \hat{\mathbf{V}}_{b,i}^s} \frac{\partial \hat{\mathbf{V}}_{b,i}^s}{\partial \hat{\mathbf{V}}_{int,i}^s} \delta \hat{\mathbf{V}}_{int,i}^s \right] 
  - \sum_{i=0}^{P} \phi w_i \left[ 
  \frac{\partial \tilde{\mathbf{f}}_{1,i}^n}{\partial \hat{\mathbf{Q}}_{b,i}^n} 
  \frac{\partial \hat{\mathbf{Q}}_{b,i}^n}{\partial \hat{\mathbf{V}}_{b,i}^n} \frac{\partial \hat{\mathbf{V}}_{b,i}^n}{\partial \hat{\mathbf{V}}_{int,i}^n} \delta \hat{\mathbf{V}}_{int,i}^n \right] \\
  + & \int_{\Omega} \nabla \phi \cdot \frac{\partial \mathbf{F}}{\partial \mathbf{Q}} \frac{\partial \mathbf{Q}}{\partial \mathbf{V}} \delta \mathbf{V}_{int} d\Omega
\end{aligned}
\end{equation}
and
\begin{equation}
\label{eq::app_B_Euler2D_DG_linearized_ref_V}
\begin{aligned}
  & RHS_{ref}\left(\hat{\mathbf{V}}, \delta \hat{\mathbf{V}}_{ref} \right) = \\
    & \sum_{i=0}^{P} \phi w_i \left[ 
  \frac{\partial \tilde{\mathbf{f}}_{1,i}^w}{\partial \hat{\mathbf{Q}}_{b,i}^w} \frac{\partial \hat{\mathbf{Q}}_{b,i}^w}{\partial \hat{\mathbf{V}}_{b,i}^w} \frac{\partial \hat{\mathbf{V}}_{b,i}^w}{\partial \hat{\mathbf{V}}_{ref,i}^w}
  \delta \hat{\mathbf{V}}_{ref,i}^w \right] 
  + \sum_{i=0}^{P} \phi w_i \left[ 
  \frac{\partial \tilde{\mathbf{f}}_{1,i}^s}{\partial \hat{\mathbf{Q}}_{b,i}^s} \frac{\partial \hat{\mathbf{Q}}_{b,i}^s}{\partial \hat{\mathbf{V}}_{b,i}^s} \frac{\partial \hat{\mathbf{V}}_{b,i}^s}{\partial \hat{\mathbf{V}}_{ref,i}^s} \delta \hat{\mathbf{V}}_{ref,i}^s \right]  \\
  - & \sum_{i=0}^{P} \phi w_i \left[ 
  \frac{\partial \tilde{\mathbf{f}}_{1,i}^n}{\partial \hat{\mathbf{Q}}_{b,i}^n} 
  \frac{\partial \hat{\mathbf{Q}}_{b,i}^n}{\partial \hat{\mathbf{V}}_{b,i}^n} \frac{\partial \hat{\mathbf{V}}_{b,i}^n}{\partial \hat{\mathbf{V}}_{ref,i}^n} \delta \hat{\mathbf{V}}_{ref,i}^n \right]
\end{aligned}
\end{equation}
In the above $RHS_{ref}$ has two less terms than $RHS_{int}$. 

\end{document}